\documentclass[twocolumn, twocolappendix]{aastex701}

\usepackage{amsmath, amssymb}
\usepackage{bm}
\usepackage{graphicx}
\usepackage{subcaption}
\usepackage{enumitem}
\usepackage{color}

\newcommand{\pII}{{\hyperlink{ac:pII}{Paper II}}}

\newcommand{\avg}[1]{\ensuremath{\mathinner{\left\langle #1 \right\rangle}}}
\newcommand{\paren}[1]{\ensuremath{\mathinner{\left( #1 \right)}}}

\newcommand{\tagb}{\ensuremath{\tau_{\rm AGB}}}
\newcommand{\vwd}{\ensuremath{v_{\rm rms}^2}}

\newcommand{\oavg}[1]{\ensuremath{\mathinner{\overline{\overline{#1}}}}}

\newcommand{\pD}[2]{\frac{\partial #2}{\partial #1}}

\newcommand{\D}[2]{\frac{{\rm d} #2}{{\rm d} #1}}

\newcommand\bb[1]{\mbox{\boldmath{$#1$}}}

\newcommand\bcdot{\,\bb{\cdot}\,}

\newcommand{\ez}{\hat{\bb{z}}}

\newcommand{\rmd}{{\rm d}}

\newcommand{\cross}{\bb{\times}}
\newcommand{\vrms}{v_{\rm rms}}
\newcommand\mn[1]{\left\langle{#1}\right\rangle}

\def\apjs{Astrophys.~J.~Supp.~Ser.}

\def\apjl{Astrophys.~J.~Lett.}
\def\aap{Astron.~Astrophys.}

\renewcommand{\eqref}[1]{Eq.~\ref{#1}}

\begin{document}

\title{White Dwarf Natal Kicks as Velocity-Space Random Walks. I. \\ Fokker-Planck Theory}
\shorttitle{Fokker-Planck Theory of White-Dwarf Natal Kicks}
\shortauthors{Majeski \& Pham et al.}

\author[orcid=0000-0002-7879-060X]{Stephen~Majeski}
\affiliation{JILA, University of Colorado and National Institute of Standards and Technology, 440 UCB, Boulder, CO 80309-0440, USA}
\email[show]{stephen.majeski@colorado.edu}  

\author[orcid=0000-0002-0924-8403]{Dang~Pham}
\affiliation{Department of Astrophysical and Planetary Sciences, University of Colorado, 391 UCB, Boulder, CO 80309-0391, USA}
\affiliation{JILA, University of Colorado and National Institute of Standards and Technology, 440 UCB, Boulder, CO 80309-0440, USA}
\email[show]{dang.pham@colorado.edu}  

\author[orcid=0000-0003-2012-5217]{Taeho~Ryu}
\affiliation{JILA, University of Colorado and National Institute of Standards and Technology, 440 UCB, Boulder, CO 80309-0440, USA}
\affiliation{Department of Astrophysical and Planetary Sciences, University of Colorado, 391 UCB, Boulder, CO 80309-0391, USA}
\email[]{Taeho.Ryu@colorado.edu} 

\author[orcid=0009-0003-6995-1840]{Gabriel Tomassini}
\affiliation{Université Côte d’Azur, Observatoire de la Côte d’Azur, CNRS, Lagrange, CS 34229, Nice, France}
\email[]{}

\author[orcid=0000-0003-3891-7554]{Andrea~Chiavassa}
\affiliation{Université Côte d’Azur, Observatoire de la Côte d’Azur, CNRS, Lagrange, CS 34229, Nice, France}
\email[]{andrea.chiavassa@oca.eu} 

\begin{abstract}
White dwarfs are the most common end state of stellar evolution, including that of our own Sun, making their formation critical to completing a more comprehensive understanding of stellar evolution and even determining the eventual fate of the solar system. Wide binaries containing white dwarfs offer a powerful probe: their components evolve largely independently, yet stellar mass loss and velocity impulses (natal kicks) on the white dwarf progenitor leave measurable imprints on orbital properties. 
Previous population studies typically assume that each white dwarf receives only a single natal kick as it forms. While convenient, this simplification conflicts with simulations finding the formation process to be episodic. Therefore, a more realistic picture is that white-dwarf progenitors experience multiple stochastic kicks over their evolution. Here, we present a Fokker-Planck model for the evolution of binary orbital parameter distributions subject to many, randomly-oriented mass-loss-driven kicks. This model recovers key features of white-dwarf binary observations, including a --2 power law tail in the probability distribution of separations, and the thermalization of binary eccentricities beyond separations of 1000 au. We additionally investigate how white-dwarf kicks impact a myriad of other processes, such as binary dissolution, planetary system evolution, and even the production of interstellar objects.
\end{abstract}

\keywords{\uat{Stellar Dynamics}{1596} --- \uat{Stellar Evolution}{1599} --- \uat{White Dwarf Stars}{1799} -- \uat{Binary Stars}{154} --- \uat{Planetary Dynamics}{2173}}

\section{Introduction}\label{sec:intro}

In their study of \textit{Gaia}'s wide binary population, \citet{ElBadry2018} discovered that a significant difference exists between the separation distributions of binaries that contain white dwarfs (WDs), and those that do not. Those authors hypothesized that the observed differences were the result of a WD natal kick --- an impulsive acceleration event that is thought to occur to each WD progenitor as it sheds the mass necessary to achieve its final, compact form.
Such natal kicks have been invoked to explain other WD observational mysteries as well. In core-collapse globular clusters, young WDs are observed to have a less concentrated radial distribution than older ones~\citep{Heyl2007}. This can be explained if, when they are born, WDs receive a sudden velocity boost over the cluster crossing timescale of $\sim 10^5$ years~\citep[e.g.,][]{Davis2008, Fregeau2009}. Natal kicks are also likely responsible for the deficit of WDs in open clusters~\citep{fellhauer2003}. Importantly, the kick strengths necessary to explain the globular cluster, open cluster, and stellar-binary observations are all roughly ${\sim} 1~\rm km/s$, lending credence to the idea that such natal kicks do indeed occur. Since \citet{ElBadry2018}, WD natal kicks have received increased attention in the context of binary stellar evolution. \citet{hwang2025} considered their role in determining the eccentricity distribution of wide binaries, evolving the model for these kicks from a sudden impulse to a gradual acceleration. \citet{oconnor2026} similarly considered gradually accelerating kicks, although they focused on how kick dynamics interact with the AGB star envelope to produce strong tidal forces and even induce common-envelope evolution. Through all of these investigations, however, the exact physics that leads to WD natal kicks has remained largely elusive.

At the most fundamental level, WD natal kicks are at least known to be related to stellar mass loss that occurs after stars leave the main sequence (MS). In fact, these kicks are referred to as ``natal'' because their origin lies in the WD formation process itself (as opposed to kicks that occur after WD formation, such as those imposed by stellar flybys). During the asymptotic giant-branch (AGB) phase of a low-to-intermediate-mass star's life, it sheds the outer layers of its atmosphere through a combination of winds and, at late times, massive shell-ejection events that accompany thermal pulses~\citep[e.g.,][]{Hofner2018, Decin2019, Decin2021, Freytag2023,maercker2024}. As a result of these processes, over the entire AGB lifetime an order-unity fraction of the star's mass is lost to its surroundings. Natal kicks are thought to be the result of asymmetries in this mass loss; if an excess of mass is ejected in one particular direction, the star will recoil in the opposite direction. This explanation appears to be supported by observations of the stellar chromospheres of red super giant and AGB stars, which have been shown to be asymmetric~\citep{Ohnaka2016, Ohnaka2017, Ohnaka2025}. A better understanding of WD natal kicks may therefore be used to infer information about these late stages of stellar evolution and the associated mass loss, which in turn can improve stellar modeling and population synthesis codes~\citep[e.g.,][]{Paxton2011,Breivik2020}.

The effects of mass loss and natal kicks on post-MS stars also have significant impacts on the dynamics of evolved planetary systems.
Planetary orbits are expected to expand due to mass loss, but natal kicks can introduce diffusion of these orbits as well~\citep[e.g.,][]{sackman1993,Tremaine2023, esseldeurs2026}.
This can seed instability in post-MS planetary systems, potentially contributing to WD pollution~\citep[e.g.,][]{Veras2016, Veras2024}, or leading to the ejection of small bodies like asteroids and comets to produce interstellar objects~\citep[e.g.,][]{Katz2018}.
Interstellar objects can even result directly from kick-induced modifications to the orbits of distant Oort-cloud-like objects \citep{OConnor2023}.

In all of the aforementioned studies of WD natal kicks, the process has been assumed to involve a single acceleration event, whether it be impulsive or gradual. However, \citet{hwang2025} stress that this is not the only possibility: the observational evidence of WD natal kicks could just as easily be interpreted as evidence that AGB stars experience many uncorrelated kicks, with the ``total'' observed kick being the final result of a random walk in velocity, rather than a single, ${\sim} 1 ~\rm km/s$ perturbation. In this paper (Paper I), which is the first of two, we investigate this possibility by exploring how the dynamics of stellar binaries are affected by many randomly-oriented recoil kicks\footnote{In our model, these mass-loss associated kicks on the TP-AGB star are assumed to be episodic, impulsive events, in contrast with the slowly varying recoil explored in \citet{oconnor2026}.} that occur as a result of stochastic, asymmetric stellar mass loss. Unlike previous studies, we assume that these kicks happen primarily during the thermally pulsing (TP) AGB phase. We do this for two main reasons: First, observational evidence from globular clusters points to these kicks occurring over a ${\sim} 10^5\mathrm{~yr}$ timescale~\citep{Heyl2007}. This is much more comparable to the TP-AGB lifetime~\citep[$\sim10^5{-}10^6$ yr;][]{Girardi2007,Kalirai2014} than the entire AGB lifetime of a typical star \citep[$\sim10^6{-}10^7$ yr;][]{siess2006a,salaris2014}. Second, both simulations and observations indicate that this phase is when most of the stellar mass loss occurs, and that this mass loss is accompanied by pulsations of the stellar structure similar in form to the kind of episodic behavior we seek to model~\citep{maercker2012,mcdonald2016,cui2026}. The main goal of this paper is therefore to construct a theoretical model of expected WD-binary separations and eccentricities in light of this stochastic and asymmetric mass loss process. 

The approach that we take to develop such a model is statistical: a Fokker-Planck equation is derived that we solve, both numerically and analytically, for the evolution of energy and eccentricity probability distributions of two-body systems. We then employ our findings to explain key features in the separation and eccentricity distributions of \textit{Gaia}'s WD binaries \citep{ElBadry2018, hwang2025}. Specifically, we demonstrate that our model reproduces a broken power law with a $-2$ tail in the wide-binary regime of the separation distribution \citep{ElBadry2018}, and that it explains the thermalization of the eccentricity distribution of WD-MS binaries at separations $\gtrsim 10^3 \mathrm{~au}$ \citep{hwang2025}. We also explore the consequences that this random walk holds for expected distributions of planets and debris around WDs.

In a companion paper (\citealt{paper2}; hereafter \hypertarget{ac:pII}{Paper II}), we focus on stellar binaries alone and present a direct comparison between a model of random-walking WD-progenitors and observational stellar-binary distributions in order to infer properties of the kicks themselves. The precision necessary to accomplish that task requires going beyond the Fokker-Planck approach taken here: We construct a binary population synthesis model where TP-AGB stars experience multiple randomly oriented kicks. With those simulations, we then directly fit our model to observational data, inferring properties like the number of kicks, the strength of the kicks, and even their dimensionality. We also explore the total rates of binary dissolution and transitions into the mass-transfer regime. For details on that investigation, we refer the reader to \pII. In this work (Paper I), the only result that we make reference to from \pII~is the inferred strength of the kicks, which itself is similar to estimates obtained by previous works. We emphasize, however, that many of the results presented here are independent of the kick strengths, so long as they occur as a random walk.

This paper is structured as follows.
In \S\ref{sec:thry}, we derive the Fokker-Planck theory for the evolution of energy and eccentricity probability distributions of stellar binaries.
In \S\ref{sec:nsln}, we numerically solve the evolution equation of \S\ref{sec:thry} to probe how planetary systems, debris disks, and stellar binaries respond to stochastic TP-AGB kicks.
In \S\ref{sec:disc}, we compare our findings with existing observational and modeling literature and discuss future directions.
In \S\ref{sec:conclusion}, we summarize the results of this paper.

\section{Fokker-Planck Theory}\label{sec:thry}

To describe the distribution of stellar binaries in this many-kick paradigm, we now develop a statistical model for the gravitational dynamics of binary systems containing TP-AGB stars. More precisely, we seek to evolve the probability distribution of orbital parameters for a large number of non-interacting binary systems where one component experiences many impulsive kicks and mass loss. 
As part of this effort, we derive equations that are easily generalizable to a TP-AGB companion of any mass (denoted by $m_{\rm C}$), so that they may be applied beyond the stellar-binary case alone. 
Accordingly, while our primary motivation and application is to the explanation of WD binary populations, we emphasize that the results obtained in this section can (and will) be applied to a broad range of two-body systems containing TP-AGB stars (e.g., non-interacting exoplanets, \S\ref{sec:planets}, and diffuse debris disks, \S\ref{sec:debris}). 

The mathematical approach that we take is to develop a Fokker-Planck equation that evolves the probability distribution ($P$) of the binary's specific orbital energies
\begin{equation}
    E = -\frac{\mu}{2a} = -\frac{\mu}{r} +\frac{v^2}{2}
\end{equation}
and squared eccentricities
\begin{equation}
    \varepsilon \equiv e^2 = 1+\frac{2EJ^2}{\mu^2}
\end{equation}
in the presence of velocity-space kicks and mass loss. In the above expressions, $a$ is the orbital semi-major axis, $e$ is the eccentricity, $\mu = G(m_{\rm AGB}+m_{\rm C})$ is the standard gravitational parameter, and $J = |\bb{J}| = |\bb{r}\cross \bb{v}|$ is the magnitude of the specific angular momentum (with $m_{\rm AGB}$ the TP-AGB mass, $\bb{r}$ the binary separation vector and $\bb{v}$ the relative velocity vector). In the Fokker-Planck framework, the evolution of $P$ is divided into contributions resulting from deterministic and random forces, the former being represented by an advection of probability in the $E{-}\varepsilon$ phase space, and the latter by diffusion. For example, a decrease in mass of the TP-AGB star exclusively increases the orbital energy (yielding advection), but randomly-directed kicks can both increase and decrease the energy (yielding diffusion). When coefficients for the rates of advection of probability ($A$) and diffusion of probability ($D$) are determined in our $E{-}\varepsilon$ phase space, the evolution of $P$ is given by
\begin{multline}
    \pD{t}{P} = -\pD{E}{}\left( A_E P\right)-\pD{\varepsilon}{}\left( A_\varepsilon P\right) + \pD{E^2}{^2}\left( D_E P\right) \\
    + \pD{\varepsilon^2}{^2}\left( D_\varepsilon P\right) + \pD{E\partial\varepsilon}{^2}\left( D_{E,\varepsilon} P\right),
\end{multline}
where the subscripts on $A$ and $D$ refer to the quantities being advected and diffused~\citep{chandrasekhar1943,risken1989}, respectively. The primary task in our Fokker-Planck approach is therefore to determine these rates.

\subsection{Jump Moments with Mass Loss}\label{sec:jmom}

Each kick, with its associated mass loss, introduces a perturbation to the orbital energy $\Delta E$ and to the squared eccentricity $\Delta \varepsilon$. In our treatment, we assume that all kicks are of an equivalent magnitude ($\Delta v$), and that the mass lost with each kick is constant (which we apply to the gravitational parameter as $\Delta \mu$).\footnote{In real TP-AGB stars, it is expected that $\Delta v$ and $\Delta \mu$ vary over the TP-AGB lifetime. Determining kick strengths and mass-loss variation for these stars remain open problems, however. In the absence of definitive prescriptions for $\Delta \mu(t)$ and $\Delta v(t)$, we make the simple assumption here that they are constant and leave more sophisticated treatments to future work.} The random nature of the kicks enters via the direction in which they impart momentum to the recoiling TP-AGB star. For example, some kicks will be directed along the orbital motion of the TP-AGB star, speeding it up and increasing its energy, while others will be against the direction of motion, slowing it down and reducing its energy. Furthermore, a kick that is aligned with orbital motion at pericenter would be misaligned with orbital motion if it instead occurred at apocenter. It is therefore impractical to describe the advection and diffusion of $P$ according to the properties of every individual kick that occurs at a specific phase of the orbital motion. Rather, the $A$ and $D$ are given in terms of $\Delta E$ and $\Delta \varepsilon$ after averaging over many independent kicks (e.g., with kick averages denoted by $\overline{\Delta E}$) and \textit{then} averaging over different binaries at unique orbital phases (with the combined kick-orbit double average denoted by $\oavg{\Delta E}$). If the kicks occur once per fixed time interval $\Delta t$, then the rates of advection are given by 
\begin{equation}\label{eq:advec}
    A_E = \frac{\oavg{\Delta E}}{\Delta t}, \; A_\varepsilon = \frac{\oavg{\Delta \varepsilon}}{\Delta t},
\end{equation}
and the rates of diffusion are given by 
\begin{equation}\label{eq:diff}
    D_E = \frac{\oavg{(\Delta E)^2}}{2\Delta t}, \; D_\varepsilon = \frac{\oavg{(\Delta \varepsilon)^2}}{2\Delta t}, \; D_{E,\varepsilon} = \frac{\oavg{\Delta E\Delta \varepsilon}}{\Delta t}.
\end{equation}
The numerators in each coefficient (e.g., $\oavg{\Delta E}$ and $\oavg{(\Delta E)^2}$ for the energy) are referred to as the advection and diffusion ``jump moments''~\citep[or alternatively the Kramers-Moyal coefficients;][]{kramers1940,moyal1949,cohn1979}.

Temporarily putting aside the effects of mass loss on the energy and eccentricity evolution, the role of velocity kicks alone in diffusing and advecting binary orbital parameters has been studied in great detail by \citet{hamilton2024b}, and more recently by \citet{caputo2026}. In \citet{hamilton2024b}, the $E$ and $\varepsilon$ jump moments were determined for kicks of arbitrary amplitude $\Delta v$, although the authors applied this work primarily to study the role of stellar flybys in wide binary dynamics. Nonetheless, the general nature of their treatment of random velocity kicks allows us to borrow from their framework --- we need only account for the additional effect of mass loss on binary orbital evolution. That said, the process of deriving the jump moments with mass loss is still cumbersome, therefore, we reserve the details of that derivation for Appendix \ref{app:jmom}, providing only a brief overview of the steps and the outcome here. 

First we address the orbital energy jump moments. The derivation, detailed in Appendix \ref{app:jmom}, begins with perturbing the energy by applying the instantaneous changes $\Delta \bb{v}$ (with a random direction) and $\Delta \mu$ to $\bb{v}$ and $\mu$ respectively. The instantaneous nature of the kicks means that $\bb{r}$ is not immediately modified. The pre-kick energy is subtracted from the perturbed energy to obtain $\Delta E$, and the result is averaged over all possible kick orientations and orbital phases to obtain the desired jump moments. The advection moment of the energy is found to be
\begin{equation}\label{eq:eajmp}
    \oavg{\Delta E} = -\frac{\Delta\mu}{a} + \frac{(\Delta v)^2}{2} = 2E\paren{\frac{\Delta\mu}{\mu}} + \frac{(\Delta v)^2}{2},
\end{equation}
and the diffusion moment is 
\begin{multline}\label{eq:edjmp}
    \oavg{(\Delta E)^2} = -2E\paren{\frac{(\Delta v)^2}{d}} + \frac{4E^2}{\sqrt{1-\varepsilon}} \paren{\frac{(\Delta \mu)^2}{\mu^2}} \\ 
    + 2\frac{E}{\mu}(\Delta v)^2\Delta \mu + \frac{(\Delta v)^4}{4}.
\end{multline}
Here, $d$ is the dimensionality of the kicks: their random direction can be restricted to a particular axis ($d=1$), a particular plane ($d=2$), or completely unrestricted ($d=3$). 

Strictly speaking, one can insert \eqref{eq:eajmp} and \eqref{eq:edjmp} as they are directly into \eqref{eq:advec} and \eqref{eq:diff}. However, additional simplification is possible if we first make use of the implied relationship that exists between $\Delta \mu$ and $\Delta v$. The velocity of the star undergoes a random walk, so the root-mean-square total kick strength\footnote{Previous single-kick studies also employ a $v_{\rm rms}$, however in those instances the $v_{\rm rms}$ sets the width of the distribution that they sample a single kick from for each star. Here, the random walk naturally produces a $v_{\rm rms}$ via the final distribution of perturbed stellar velocities, even though we assume fixed $\Delta v$.} after $N$ kicks is given by $v_{\rm rms}^2 = N(\Delta v)^2$. On the other hand, the assumed mass loss is linear, so it obeys the relationship $N\Delta \mu = -f\mu$, where $f$ is the total fractional change to the gravitational parameter \citep[which itself is $\mathcal{O}(1)$;][]{Vassiliadis1993,Vassiliadis1994}. $\Delta v$ and $\Delta \mu$ are therefore related through the total number of kicks/mass-loss events that a TP-AGB star experiences. If many kicks occur ($N\gg1$) such that the time interval between these kicks becomes small relative to the TP-AGB lifetime ($\Delta t \ll \tau_{\rm AGB}$, with $\tagb$ the TP-AGB lifetime), then an asymptotic ordering may be constructed based on the largeness of $N$, and its relationships to $\Delta v$ and $\Delta \mu$:
\begin{equation}\label{eq:ord}
    \frac{|\Delta \mu|}{\mu} \sim \frac{(\Delta v)^2}{v_{\rm rms}^2} \sim \frac{(\Delta v)^2}{|E|} \sim \frac{1}{N} \equiv \epsilon \ll 1, 
\end{equation} 
where we have introduced the expansion parameter $\epsilon$. By including the third term of this ordering, we have made the asymptotic assumption that $v_{\rm rms}^2 \sim |E|$. Importantly, we are not approximately equating the RMS kick energy to the orbital energy, rather, we are assuming that the ratio of these two quantities satisfies $N^{-1} \ll v_{\rm rms}^2/|E| \ll N$. Given \eqref{eq:ord}, we divide \eqref{eq:eajmp} and \eqref{eq:edjmp} by $\Delta t=\tau_{\rm AGB}/N$, and keep only terms that are of order $E/\tagb \times \epsilon^0$ for the advection moment and $E^2/\tagb \times \epsilon^0$ for the diffusion moment. We find that the leading-order advection coefficient in energy is
\begin{equation}\label{eq:enadv}
    A_E = \frac{\oavg{\Delta E}}{\Delta t} \approx \frac{1}{2}\frac{v_{\rm rms}^2}{\tau_{\rm AGB}} - 2\frac{fE}{\tau_{\rm AGB}},
\end{equation}
and the diffusion coefficient is\footnote{The exclusion of the second term in \eqref{eq:edjmp} implies that we ignore systems with eccentricities sufficiently close to 1 that the ordering in $N$ is violated (i.e., we assume that the factor $\sqrt{1-\varepsilon}$ is not $\mathcal{O}(\epsilon)$). For $N\gtrsim 100$, this only becomes a problem for eccentricities $e\gtrsim 1-10^{-5}$.}
\begin{equation}\label{eq:endiff}
    D_E = \frac{1}{2}\frac{\oavg{\Delta E^2}}{\Delta t} \approx -\frac{1}{d}\frac{Ev_{\rm rms}^2}{\tau_{\rm AGB}}.
\end{equation}
When $d=3$ is assumed and $\Delta t$ is normalized by $\tagb$, these advection and diffusion coefficients recover exactly the result of \citet{hamilton2024b}, save for the addition of a new mass-loss term in the advection coefficient.

The process that we applied to the energy to obtain those jump moments can similarly be applied to the angular momentum for the $\varepsilon$ moments. The result, as obtained in Appendix \ref{app:jmom}, is
\begin{multline}
    \oavg{\Delta\varepsilon} = (\Delta v)^2 \paren{\frac{-2-d+\varepsilon(7+d)}{2 Ed}} \\
    + (\Delta v)^2\frac{\Delta\mu}{\mu}  \frac{(1-\varepsilon)}{E} + \frac{(\Delta\mu)^2}{\mu^2}(1-\varepsilon) +\mathcal{O}\left(\frac{\varepsilon}{N^3}\right)
\end{multline}
for the advection moment of $\varepsilon$ and
\begin{multline}
        \oavg{(\Delta\varepsilon)^2} = - (\Delta v)^2\frac{5 \varepsilon (1-\varepsilon)}{E d} \\
    - \frac{(\Delta v)^2}{E}\paren{\frac{\Delta\mu}{\mu}}(1-\varepsilon)^2\paren{2+\frac{2\varepsilon}{d(\varepsilon-1)}} +\mathcal{O}\left(\frac{\varepsilon^2}{N^3}\right)
\end{multline}
for the diffusion moment. For brevity, we have not explicitly written out terms beyond order $N^{-2}$ in each expression, as they are neglected in the ordering step anyway. Employing \eqref{eq:ord} to reduce the leading-order advection and diffusion coefficients that follow from these $\varepsilon$ jump moments, we find that
\begin{equation}\label{eq:epadv}
    A_\varepsilon = \frac{\oavg{\Delta \varepsilon}}{\Delta t} \approx \vwd\left(\frac{ (7+d)\varepsilon-2-d}{2dE\tagb}\right),
\end{equation}
and
\begin{equation}\label{eq:epdiff}
    D_\varepsilon = \frac{1}{2}\frac{\oavg{\Delta \varepsilon^2}}{\Delta t} \approx \frac{5}{d}\left(\frac{\vwd (\varepsilon^2-\varepsilon)}{2E\tagb}\right).
\end{equation}
For $d=3$, these coefficients are exactly those obtained by \citet{hamilton2024b}, with no change resulting from the mass loss at leading order. This is expected, as slow mass loss has no effect on the eccentricity of Keplerian orbits~\citep{jeans1924}.

The only remaining moment to be determined is that of the cross-diffusion term $D_{E,\varepsilon}$. We simply note here that it has no contribution at order $\epsilon^0$, and so it is neglected for the remainder of this work (see Appendix \ref{app:jmom} for details).

\subsection{The $E{-}\varepsilon$ Fokker-Planck Equation}\label{sec:fpeq}

With the advection and diffusion coefficients known, we construct the Fokker-Planck equation for the evolution of $P(E,\varepsilon)$, which, following minor manipulations, is:
\begin{multline}\label{eq:fp}
    \frac{\partial P}{\partial t} = \frac{1}{\tagb} \left(2fE-\frac{d+4}{2d}\vwd\right)\frac{\partial P}{\partial E}  + \frac{2f}{\tagb}P\\
    - \frac{E}{d}\left(\frac{\vwd}{\tagb}\right)\frac{\partial^2 P}{\partial E^2} - \vwd\pD{\varepsilon}{}\left( \frac{(d-3)\varepsilon+3-d}{2dE\tagb}P\right) \\
    + \frac{\partial}{\partial \varepsilon} \left( \frac{5\vwd (\varepsilon^2-\varepsilon)}{2dE\tagb} \frac{\partial P}{\partial \varepsilon}\right).
\end{multline}
We remind the reader that this equation assumes that binary orbits are Keplerian, that individual kicks are impulsive, that many kicks occur ($N\gg 1$), and that the ratio $|E|/\vwd$ is both $\ll N$ and $\gg N^{-1}$. 

There are many competing physical processes present in \eqref{eq:fp}. The energy distribution, for example, is shaped by a competition between drift due to mass loss, drift from the net energy imparted by the kicks, and diffusion from the kicks. Importantly, the energy distribution is not affected by the eccentricity evolution, a consequence of the assumption $N\gg1$ (which allowed us to neglect the $\varepsilon$-dependent term in \eqref{eq:edjmp}). On the other hand, the eccentricity distribution is dependent on that of the energy, with the diffusion and advection coefficients growing as $E\rightarrow 0$. Additionally, the eccentricity advection term, as given in \eqref{eq:fp}, vanishes entirely for $d=3$ \citep[in agreement with][]{magorrian1999, hamilton2024b}. The presence of so many distinct processes that vary with mass-loss and kick parameters implies that there are regimes where certain terms in \eqref{eq:fp} dominate over others.

\begin{figure*}[ht]
\centering
\includegraphics[width=0.85\linewidth]{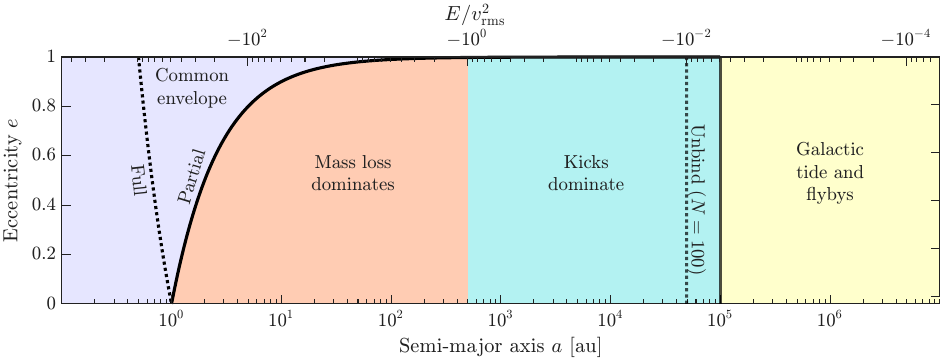}
\caption{Diagram of the regions of $a{-}e$ phase space in which different physics from \eqref{eq:fp} is expected to be dominant. The relationship between $E$ and $a$ assumes $m_{\rm total} = 1~\rm M_\odot$ and an RMS total kick strength of 1 km/s. In the region interior to the solid lines, we expect solutions of \eqref{eq:fp} to be largely unaffected by extraneous physics.}
\label{fig:diag}
\end{figure*}
In Fig. \ref{fig:diag}, we illustrate the different physical regimes present in the phase space of \eqref{eq:fp}, translated into the orbital parameters $a$ and $e$. We furthermore compare the physics in \eqref{eq:fp} to effects that, although extraneous to our model, are expected to act as boundary conditions to the dynamics captured in our Fokker-Planck treatment. Beginning with the physics that we \textit{do} account for in \eqref{eq:fp}, we divide the semi-major-axis phase space (a proxy for energy) into two regions depending on whether kicking or mass loss is the dominant effect in the evolution of $P$. The border between these regions is the semi-major axis at which $\vwd = E$, the numerical value of which is ${\approx}500$ au (assuming $m_{\rm AGB}+m_{\rm C} = 1~\rm M_\odot$ to relate $E$ to $a$, and $v_{\rm rms} = 1$ km/s for the strength of the kicks).\footnote{The RMS kick strength assumed here is informed by the result that we obtain in \pII~by comparing the random-walk model with binary observational data. It is also roughly consistent with the results of earlier studies~\citep[e.g.,][]{Heyl2007,ElBadry2018}.} For $a$ larger than this value, kicks dominate advection and kick-induced diffusion becomes important for the evolution of $P$ (in both energy and eccentricity). For $a$ less than this value, mass loss dominates evolution on its own. Mass loss dominates at low $a$ because orbital velocities increase as binaries become more bound, meaning a fixed $v_{\rm rms}^2$ is a smaller fraction of the orbital energy at smaller $a$ (whereas a fractional change in $\mu$ always changes $E$ by the same fraction, regardless of separation). At high separations, the $\vwd/E$ increases while $f$ remains unchanged, so the balance between mass loss and kicks inevitably reverses in favor of the kicks at sufficiently large $a$. 

Of course, realistic binaries are also subject to external forces that we do not account for in our Fokker-Planck model. For example, perturbations from the Galactic tidal force and stellar flybys become sufficiently strong to unbind binaries at semi-major axes exceeding $\sim 10^5$ au~\citep{Heisler1986,jiang2010,modak2023,hamilton2024b}. That said, the timescales on which these forces operate far exceed the TP-AGB lifetime $\tagb$: Below $a\sim10^5 \; \rm au$, flybys and the Galactic tide require $>10^8 \rm \; yr$ to meaningfully modify the orbital-parameter distribution of binaries~\citep{hamilton2024b}. We therefore consider the main consequence of these forces to be the dissolution of binaries that are pushed beyond $a\approx10^5 \rm \; au$ in the long-time limit (post-AGB evolution).

In the opposite limit where the binary separation is so small that it approaches the radius of the TP-AGB envelope \citep[$\sim 1 \rm \; au$;][]{Hofner2018}, one must consider common envelope (CE) evolution, which, unlike tides and flybys, affects orbital parameters on a timescale much shorter than $\tagb$~\citep{Ivanova2013}. We include two varieties of CE evolution in Fig. \ref{fig:diag}: full CE, where the apocenter is located within the TP-AGB star (which itself has a radius of roughly 1 au), and partial CE, where the pericenter is located inside the TP-AGB star but the apocenter is not. CE evolution rapidly damps orbital energy in binary systems, therefore, in what follows we avoid overlap with the CE regime by restricting ourselves to binaries at $a\gtrsim 10$ au. 

Lastly, although our Fokker-Planck equation is derived in the asymptotic limit of many kicks ($N\gg 1$), implying infinitesimal $\Delta v$, the exact number of kicks that a real TP-AGB star experiences is finite, and therefore implies a finite kick strength $\Delta v$ as well. For a given $v_{\rm rms}$ and $N$, there exists, then, a semi-major axis where one single kick is sufficient to unbind a given binary system. For $N=100$, $m_{\rm AGB}+m_{\rm C} = 1~\rm M_\odot$, and $v_{\rm rms}=1$ km/s, this semi-major axis is given as a vertical dotted line in Fig. \ref{fig:diag}. Even for a modest $N=100$, this value of $a$ is very nearly in the regime where binaries are already dissolved by the Galactic tidal force and stellar flybys in the long term. We therefore do not take any special measures here to account for this finite-$N$ effect (although \pII~addresses finite $N$ in detail).

\subsection{Solution for the Reduced Energy Distribution}\label{sec:fsln}

\begin{figure}[h!]
\centering
\includegraphics[width=1.0\linewidth]{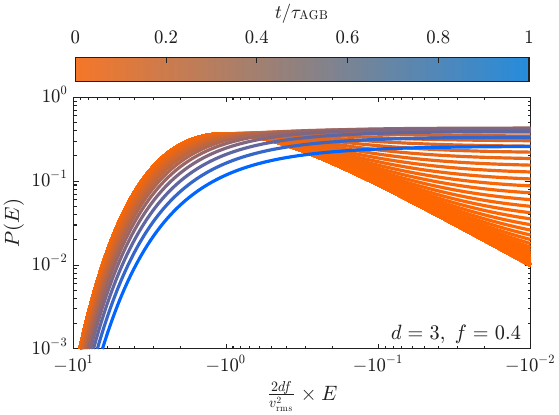}
\caption{An \textit{exact}, time-evolving solution (\eqref{eq:fpesln}) of the energy Fokker-Planck equation (\eqref{eq:fpe}). The initial distribution is given by $P_0(E) = \frac{2df}{\vwd}E\exp \left(\frac{2df}{\vwd}E\right)$, assuming $f=0.4$ and $d=3$.}
\label{fig:exdist}
\end{figure}

As mentioned in \S\ref{sec:fpeq}, the energy evolution of $P$ can be studied independently of the eccentricity dynamics. Indeed, by integrating \eqref{eq:fp} over $\varepsilon$ and assuming appropriate boundary conditions, the $\varepsilon$ dependence vanishes altogether and we are left with an equation for the reduced energy distribution $P(E)$:
\begin{multline}\label{eq:fpe}
    \frac{\partial P}{\partial t} = \frac{1}{\tagb} \left(2fE-\frac{d+4}{2d}\vwd\right)\frac{\partial P}{\partial E} \\
    - \frac{E}{d}\left(\frac{\vwd}{\tagb}\right)\frac{\partial^2 P}{\partial E^2} + \frac{2f}{\tagb}P.
\end{multline}
This equation has a solution in terms of the generalized Laguerre functions, $L_n^\nu$, given by
\begin{multline}\label{eq:fpesln}
    P(E,t) = \exp \left(\frac{2df}{\vwd}E\right)\sum_{n=0}^\infty a_{n,0} \times \\ \exp\left( -2(n+\nu)f\frac{t}{\tagb}\right)L_n^\nu\left(-\frac{2df}{\vwd}E\right).
\end{multline}
Here, $\nu = 1+d/2$, and the $a_{n,0}$ coefficients satisfy the following decomposition of the initial energy distribution:
\begin{equation}
    P_0(E) = \exp\left(\frac{2df}{\vwd}E \right) \sum_{n=0}^\infty a_{n,0} L_n^\nu\left(-\frac{2df}{\vwd}E\right).
\end{equation}
The $a_{n,0}$ can therefore be determined by exploiting the orthogonality of the Laguerre functions for integer and half-integer values of $\nu$. Although the sum in \eqref{eq:fpesln} extends to infinity, the fact that high $n$ contributions decay the fastest means that the final solution can often be approximated well by just a few $a_{n,0}$. For many (realistic) initial energy distributions relevant to stellar binaries, planets, or debris, the Laguerre decomposition is difficult to evaluate analytically, but can easily be evaluated numerically. For illustrative purposes, however, we now consider a simple initial distribution, $P_0(E) = \frac{2df}{\vwd}E\exp \left(\frac{2df}{\vwd}E\right)$, which admits the convenient decomposition $a_{n,0} = -(2+d/2)\delta_{n,0}+\delta_{n,1}$. The evolution of the solution (\eqref{eq:fpesln}) given this initial condition is shown in Fig. \ref{fig:exdist} for $d=3$ and $f=0.4$.

The evolution of the distribution in Fig. \ref{fig:exdist} can be broken down into three main features: a shift of the overall distribution to higher $E$, a flattening of the distribution at high $E$, and a decrease in the overall normalization. 

The rightward shift in the distribution is simply the consequence of advection due to mass loss and kicks (see \eqref{eq:enadv}). However, the flattening in the distribution is mediated by diffusion (\eqref{eq:endiff}). The mathematical source of this behavior can be seen in the fact that the highest $n$ Laguerre functions decay the fastest in \eqref{eq:fpesln}. At late times, then, the solution will converge toward $P(E)\propto\exp\left(\frac{2df}{\vwd}E \right) L_0^\nu\left(-\frac{2df}{\vwd}E\right)$. Noting that $L_0^\nu = 1$, in the limit of $E\rightarrow 0$ the leading order behavior for such a distribution is a constant. Looking at this result from a physical perspective, the relatively strong diffusion that is induced by kicks in the weakly bound limit causes the binaries to spread all over the energy phase space. Eventually, such diffusion will lead to a flat (or nearly flat) distribution in energy. 

When renormalized in terms of the semi-major axis, a $P(E)\propto \rm const$ energy distribution becomes $P(a)\propto a^{-2}$. \textit{This leads to one of the most significant results of this work:} 
In the kick-dominated regime, which begins around $a\sim10^3$ au (Fig. \ref{fig:diag}), we expect WD binary distributions to feature an $a^{-2}$ power law tail.
Our framework therefore explains a prominent feature of the WD binary distributions presented in \citet{ElBadry2018}: intrinsic separation ($s$) distributions for binaries containing WDs showed $s^{-2}$ power laws beyond $\log(s/{\rm au})\approx 3$. Although our result is for the semi-major axis, \citet{ElBadry2018} note that intrinsic separation and semi-major-axis distributions are likely to be very similar.\footnote{Similar power-law tails also appear to be a consequence of diffusion due to stellar flybys~\citep{weinberg1987,jiang2010,tian2020,hamilton2024b}, although this process occurs on a much longer timescale than TP-AGB evolution~\citep[see Fig. 1 of][]{hamilton2024b}, and is not likely to explain the difference between binaries containing WDs and binaries that do not. Nonetheless, flybys should at least act to preserve the power law set by TP-AGB kicks, when they do not dissolve the binaries entirely.}  This same outcome is not achieved by applying one instantaneous kick to the TP-AGB star. Our result therefore serves as strong evidence that TP-AGB stars undergo a random walk due to many small kicks in velocity space. 

The last main feature of the solution in Fig. \ref{fig:exdist} is that the normalization of the distribution decreases over time. Physically, this is because some of the binaries are dissolved by the kicks as a result of being pushed to $E>0$. That said, the Laguerre function decomposition that we employ is agnostic of the boundary condition at $E=0$, essentially allowing binaries to transit both ways across $E=0$, i.e., unbinding and re-binding. Although this treatment is not necessarily perfect, it is not entirely without physical motivation. Because orbits at such large distances evolve very slowly, a binary may be unbound by a particular kick and then re-bound by the subsequent kick if it remains sufficiently close to $E=0$. Many binaries, however, should be permanently unbound as they transit far into $E>0$, and thus should be removed from the solution. A more physical boundary condition may therefore be to progressively damp the distribution function $P(E)$ to 0 as $E$ becomes increasingly positive. For a sufficiently gradual damping, however, we expect such a change in boundary conditions to have little impact on our main results. We also note that the solution \eqref{eq:fpesln} imposes no boundary condition on the tightly bound side, so there is no implied loss here through collisions or CE evolution. 

\subsection{Thermalizing the Eccentricity Distribution}\label{sec:fecc}

The full, two-dimensional distribution $P(E,\varepsilon)$ does not admit a similarly convenient analytic solution to the reduced energy distribution $P(E)$. That said, there are useful manipulations of \eqref{eq:fp} that reveal details of the eccentricity evolution, especially concerning the thermalization of binary eccentricity distributions. In this context, a thermalized eccentricity distribution is one that is uniform in $\varepsilon=e^2$, or equivalently $P(e)\propto e$. When binary bodies experience random velocity kicks, such as in our case or with flybys, it is expected that the eccentricity distribution should approach a thermal one~\citep[cf.][]{Jeans1919, Ambartsumian1937,hamilton2024b}. Here, we prove that thermalization is indeed an outcome of TP-AGB kicks as well, and discuss how that result is relevant to the eccentricity distributions of WD-containing binaries.

Taking $d=3$ as the most likely dimensionality of the kicks, we first exchange $\varepsilon$ in favor of $\xi \equiv 1-2\varepsilon$, which leads to the following modified Fokker-Planck equation:
\begin{multline}
    \frac{\partial P}{\partial t} = \frac{1}{\tagb} \left(2fE-\frac{7}{6}\vwd\right)\frac{\partial P}{\partial E}  + \frac{2f}{\tagb}P\\
    - \frac{E}{3}\left(\frac{\vwd}{\tagb}\right)\frac{\partial^2 P}{\partial E^2}
    + \frac{5\vwd}{6E\tagb}\frac{\partial}{\partial \xi} \left( (\xi^2-1) \frac{\partial P}{\partial \xi}\right).
\end{multline}
In this form, the $\xi$ diffusion term is the Legendre differential operator. It is therefore prudent to decompose $P$ into Legendre polynomials ($p_l$) as $P(E,\xi(\varepsilon)) = \sum_{l=0}^\infty F_l(E)p_l(\xi)$, which yields a simpler evolution equation for the expansion coefficients $F_l(E)$:
\begin{multline}\label{eq:lpoly}
    \frac{\partial F_l}{\partial t} = \frac{1}{\tagb} \left(2fE-\frac{7}{6}\vwd\right)\frac{\partial F_l}{\partial E}  + \frac{2f}{\tagb}F_l\\
    - \frac{E}{3}\left(\frac{\vwd}{\tagb}\right)\frac{\partial^2 F_l}{\partial E^2}
    - \frac{5\vwd}{6E\tagb}(l+1)l\,F_l.
\end{multline}
This equation dictates that higher $l$ Legendre coefficients should decay increasingly fast, as with $n$ for the Laguerre functions. In this case, however, the $l=0$ coefficient does \textit{not} decay in time (unlike with $n$), and the decay rate of $l>0$ coefficients goes as $l^2$, rather than $l$. Thus, for $\vwd \gtrsim E$, the eccentricity distribution post-TP-AGB evolution should be described by $p_0(\xi)$, which is constant in $\xi$, and therefore also in $\varepsilon$ (i.e., thermal). This is essentially the same result achieved by \citet{hamilton2024b} for stellar flybys, although their approach showed this by investigating Casimir invariants of the non-thermal part of the distribution function. Regardless, we therefore obtain, perhaps unsurprisingly, that the semi-major-axis at which the eccentricity distribution of binaries begins to thermalize is roughly the same as that where kicks begin to dominate the energy evolution, ${\sim}10^3$ au for a solar-mass TP-AGB star. Comparing this outcome with the observational work of \citet{hwang2025}, those authors found that WD binary eccentricity distributions become thermalized somewhere between $a=10^{2.5}$ and $10^{3.5}$ au. Our random-walk framework is thus also consistent with observed WD-binary eccentricity statistics.

\section{Numerical Fokker-Planck Solutions}\label{sec:nsln}

To extend the analytic discussions of \S\ref{sec:fsln} and \S\ref{sec:fecc} to more complex and varied initial probability distributions, we now present numerical solutions to \eqref{eq:fp}. The numerical technique we employ for solving \eqref{eq:fp} in the full $E{-}\varepsilon$ phase space is based on the Legendre-polynomial decomposition (\eqref{eq:lpoly}), and detailed in Appendix \ref{app:nums}. With this numerical approach, we consider initial orbital distributions for the TP-AGB companions that approximate planetary systems (\S\ref{sec:slr}), debris disks (\S\ref{sec:debris}), and the primary motivation for this work: stellar binaries (\S\ref{sec:bnry}). In each scenario, we study the kick-induced changes and discuss the implied observational consequences that follow from TP-AGB evolution. For all of the solutions presented in this section, we assume isotropically directed kicks, i.e. $d=3$. In addition, we often require the total amount of mass loss during the TP-AGB phase. To calculate this quantity, we use the pre-computed stellar-evolution grid MIST \citep{Dotter2016}, recording the star's mass at the beginning of the TP-AGB phase and its mass at the beginning of the WD phase.\footnote{Although some significant progress has been made recently to understand the WD's initial-final mass ratio \citep[][]{Miller2026a,Miller2026b}, we emphasize that constraining post-main sequence mass loss remains an open problem. Hence, the value we use from MIST is expected to be a rough estimate.}

\subsection{Test-particle Evolution of Planetary Systems}\label{sec:slr}

In the limit where the companion's mass is much smaller than that of the evolving TP-AGB star, the above formalism can be used to describe the evolution of planetary-system bodies. That is, we can take $\mu \approx G m_\mathrm{AGB}$, since $m_\mathrm{C} \ll m_\mathrm{AGB}$, which also implies that $f$ is the fraction of mass lost by TP-AGB star alone. 

In \S\ref{sec:planets}, we study the orbital evolution of low-eccentricity distributions over a large range of semi-major axes. There are two motivations to begin by applying our Fokker-Planck model to this case. First, of course, is the desire to understand how exoplanet-like systems evolve due to TP-AGB kicks, and how they might exist around WDs. Second, however, is that it will enable us to verify the conclusions reached in \S\ref{sec:thry} through a controlled study of simple initial distributions. In \S\ref{sec:debris}, we study the evolution of debris disk-like distributions, where eccentricities possess both a higher mean and spread than the planetary case. In particular, we apply our model to study the evolution of analogs to the solar system's Kuiper belt, detached objects, and inner Oort cloud. For the inner Oort cloud, we pay particular attention to those objects that are pushed to large $a$ ($>10^5$ au) and small perihelia ($<10$ au) by TP-AGB kicks. The former may be easily removed from the system due to stellar flybys, ending up as interstellar objects, while the latter may contribute to the pollution of WDs (discussed in \S\ref{sec:future}).

Throughout this section, we allow the star to lose $\approx35.9\%$ of its mass during the TP-AGB phase, appropriate for a solar mass star~\citep[which we assume uniformly for $m_{\rm AGB}$ in this section;][]{Dotter2016}. Given the corresponding $\tagb\approx1$ Myr, this translates into a mass-loss rate of $3.6\times10^{-7} \rm \; M_\odot/yr$, consistent with estimates from theory and observation present in the literature~\citep[e.g.,][]{Vassiliadis1993,prager2022}. We also assume a total root-mean-square kick strength of 1 km/s, similar to that of previous works and the value that we obtain in \pII.

\subsubsection{Near-circular Initial Distributions}\label{sec:planets}

In this subsection, we consider the evolution of bodies with an initial eccentricity distribution given by a Rayleigh distribution\footnote{Note that here and elsewhere in the paper, we only consider bound orbits. Therefore, the eccentricity $e$ is truncated between $[0, 1)$, and the distribution is renormalized appropriately.}
\begin{equation}
    e \sim \mathrm{Rayleigh}(\sigma_{\rm R}=0.02)
\end{equation}
where $\sim$ denotes that the eccentricities obey the specified distribution, and $\sigma_{\rm R}$ is the spread of the distribution (``R'', for Rayleigh, being the distribution type).
The mean eccentricity is then $\avg{e} = \sigma_{\rm R} \sqrt{\pi/2} \approx 0.025$, where $\mn{}$ denotes an average of the probability distribution $P$. Although this value is chosen arbitrarily, it is comparable to typical eccentricities of solar system planets.\footnote{We note that exoplanets typically have higher eccentricities than solar system's values \citep[e.g.,][]{Limbach2015, Stevenson2025}, which can be attributed to biases or various formation scenarios or the diversity of dynamical mechanisms.}
In addition to planets, such an eccentricity distribution could also describe ``cold'' debris disks.
For example, the primordial Kuiper Belt is constrained to be much colder than its present-day state \citep{Stern1995}, and the debris belt around HD 121617 is constrained to have eccentricity at ${\approx} 0.03$ \citep{Perrot2023,Lovell2026}.

In all cases for this section, the initial semi-major axis distribution is given by
\begin{equation}\label{eq:normp}
    \log a/{\rm au} \sim \mathcal{N}(\mu_{\rm N}=\alpha, \sigma_{\rm N}=0.1).
\end{equation}
Here, the (base-10) logarithm of the initial semi-major axis follows a normal distribution denoted by $\mathcal{N}$ with $\mu_{\rm N}$ and $\sigma_{\rm N}$ describing the mean and standard deviation, respectively (note the subscript, for ``Normal'', which distinguishes $\mu_{\rm N}$, the normal distribution mean, from the gravitational parameter $\mu$).
The parameter $\alpha$ is used to survey a range of initial peak semi-major axes for these near-circular distributions. Our survey covers $\alpha \in [1,3.5]$, which corresponds to distributions having peak semi-major axes $a_0 \in [10,3\times10^3] \rm \; au$, where the maximum $\alpha$ is chosen to approximate the inner boundary of the solar-system Oort cloud.

\begin{figure}[htbp]
\centering
\includegraphics[width=0.95\linewidth]{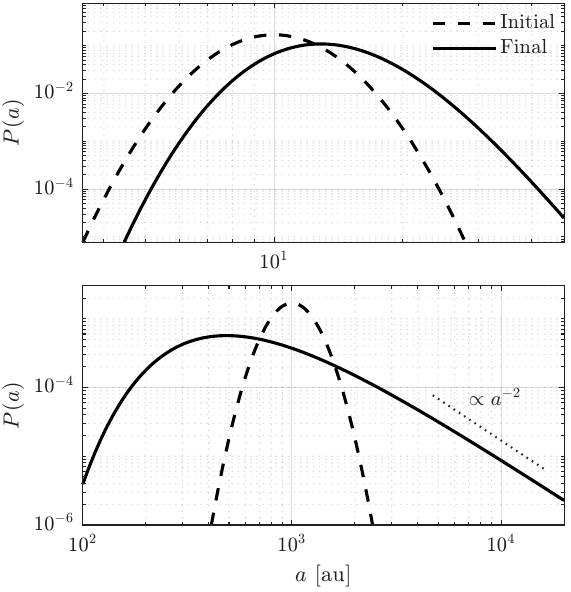}
\caption{\textbf{Top}: Pre-TP-AGB (dashed) and post-TP-AGB (solid) distributions of semi-major axes for an initially planet-like population of test particles. The eccentricity distribution was initialized as a Rayleigh distribution with a mean of $\mn{e}\approx 0.025$. The mass loss fraction is 0.359, consistent with a solar mass star, and the RMS total kick speed is 1 km/s, comparable to results from previous works~\citet{ElBadry2018,hwang2025,oconnor2026}. We find the distribution evolves adiabatically, as expected. \textbf{Bottom}: Same as the top panel, but the initial distribution is now peaked at $10^3$ au. Here, the distribution spreads diffusively, attaining an $a^{-2}$ tail as predicted in the large $a$ limit.}
\label{fig:padists}
\end{figure}
The case of $\alpha=1$, for example, which implies $a_0 = 10$ au, corresponds most closely to the evolution of planets.
Furthermore, ALMA surveys find that the typical dust disk size is also a few tens of au \citep[e.g.,][]{Zhang2025}. We show the evolution of such a distribution from its pre-kicks (dashed) to post-kicks (solid) state in the top panel of Fig. \ref{fig:padists}. In this case, the evolution is limited to a shift of the semi-major axis toward higher values. As we discuss in \S\ref{sec:thry} this is expected from dynamical theory, since at this location, the mass loss timescale is much greater than the orbital timescale, and the orbital velocity is much greater than the kick velocity. Thus, the evolution is adiabatic and we expect the orbits to expand slowly as they respond to mass loss \citep[cf.][]{Tremaine2023}.

At large separations, adiabatic evolution no longer holds. In the bottom panel of Fig. \ref{fig:padists}, we consider the same distribution of objects (\eqref{eq:normp}), but with $\alpha=3$, corresponding to $a_0=10^3$ au. We note that circular orbits at this distance are unlikely to correspond to a physical reservoir of planetary bodies, since typical protoplanetary disk sizes are much smaller than this \citep{Zhang2025}.
Nonetheless, it is important to analyze the transition regime between adiabatic and diffusive evolution. Immediately, we see that the change in the distribution is no longer described by a simple outward drift, but rather a spread on both sides due to diffusion. In addition, when $a \gtrsim 3~000$ au the distribution attains a power-law tail $\propto a^{-2}$, as predicted in \S\ref{sec:thry}.

\begin{figure}[htbp]
\centering
\includegraphics[width=1.0\linewidth]{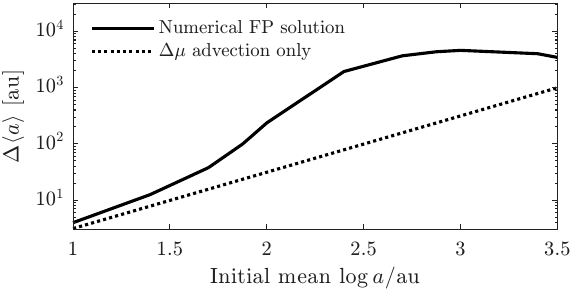}
\caption{TP-AGB-induced change in the mean semi-major axis of planet-like distributions as a function of the initial mean $\log a/\rm au$ (${=}\alpha$). The solid line denotes the value obtained from the numerical solutions of \eqref{eq:fp}, while the dotted line indicates the solution obtained by including only mass-loss-induced drift. At low initial $a$, drift from mass loss is dominant. At large $a$, however, kick-mediated drift and diffusion governs the change in $\mn{a}$.}
\label{fig:paparams}
\end{figure}
We can probe the dominant evolution process over a range of separations by considering how the mean of the semi-major axis of a given population changes ($\Delta \mn{a}$) as a function of $\alpha = \mn{\log a/\rm au}_0$. In Fig. \ref{fig:paparams}, we show this change obtained from numerical solutions to \eqref{eq:fp} (solid lines). We also show the $\Delta \mn{a}$ for solutions that are obtained by only including energy advection due to mass loss (dotted, with no diffusion or kick-induced advection). At low $\alpha$, both solutions agree well because the effects of kicks are weak compared to those of mass loss. As a result, below $\alpha \approx 2$ the orbits only expand adiabatically. As $\alpha$ grows beyond $2$, however, kicks begin to dominate the advection, which directly modifies $\mn{a}$. Additionally, the dependence of diffusion on $a$ introduces asymmetry in the shape of the distribution (cf. Fig. \ref{fig:exdist}) which further modifies $\mn{a}$, leading to an eventual plateau in $\Delta\mn{a}$.

\begin{figure}[htbp]
\centering
\includegraphics[width=1.0\linewidth]{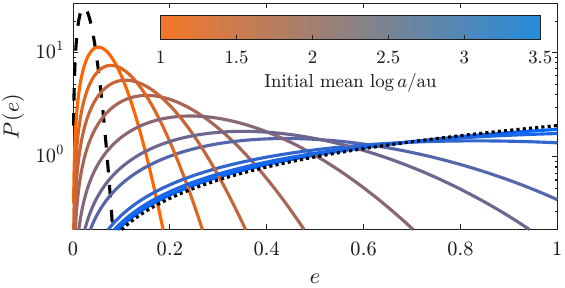}
\caption{The initial (dashed) and final (solid) eccentricity distributions for each planet-like population, as a function of initial mean $\log a/\rm au$ (${=}\alpha$). The thermal distribution is shown as a dotted line. The initial eccentricities obey a Rayleigh distribution, which is partially preserved at low initial $a$ by the final distributions. At large $a$, however, the eccentricity distributions converge to a thermal one.
}
\label{fig:pedist}
\end{figure}
In Fig. \ref{fig:pedist}, we show the corresponding initial (dashed) and final (solid) eccentricity distributions for each of the $\alpha$ given in Fig. \ref{fig:paparams}. At small distances, the eccentricity distribution spreads, causing bodies to adopt a wider range of values, but it remains similar in form to the initial Rayleigh distribution.
As we reach greater distances, the eccentricity evolution's spread gradually grows, until approximately $10^3$ au.
Beyond $\gtrsim 10^3$ au, the shape of the eccentricity distribution changes significantly and it approaches the thermal distribution (dotted line).
The increase in eccentricity spread and thermalization at greater distances have important consequences for the stability and survival of post-main sequence planetary systems \citep[e.g., see review by][]{Veras2016}.
For example, a greater spread in $e$ can destabilize planetary systems, which enables the ejection of bodies to produce interstellar objects and free-floating planets.

\subsubsection{Debris-like Initial Distributions}\label{sec:debris}

\begin{figure}[htbp]
\centering
\includegraphics[width=1.0\linewidth]{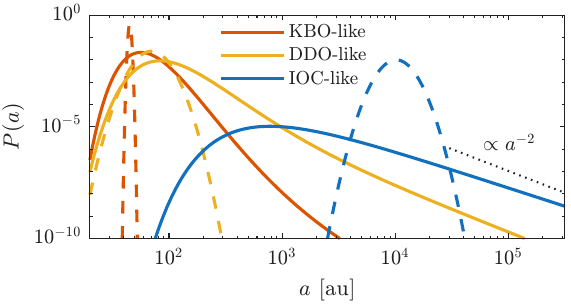}
\caption{Initial (dashed) and final (solid) semi-major axis distributions for Kuiper-belt-like, detached-disk-like, and inner-Oort-cloud-like populations of test particles around an $M_\odot$-mass TP-AGB star (assuming $v_{\rm rms}=1$ km/s kick strength). Power-law tail formation is evident in the DDO- and IOC-like populations beyond ${\sim}10^3$ au.}
\label{fig:dbadist}
\end{figure}

\begin{figure}[htbp]
\centering
\includegraphics[width=1.0\linewidth]{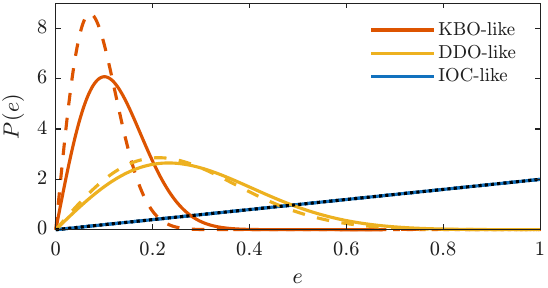}
\caption{Initial (dashed) and final (solid) eccentricity distributions for Kuiper-belt-like, detached-disk-like, and inner-Oort-cloud-like test-particle populations. The black dotted line is a thermal distribution. DDO and KBO populations were initialized as Rayleigh distributions, while the IOC population is initially thermal (hence why it does not evolve).}
\label{fig:dbedist}
\end{figure}

We now seek numerical Fokker-Planck solutions given the following debris-like initial conditions. First, we consider a Kuiper-belt-like population (abbreviated as KBO-like) with the following semi-major axis and eccentricity distributions
\begin{align}
    \log a/\rm au &\sim \mathcal{N}(\mu_{\rm N}=1.6, \sigma_{\rm N}=0.01) \\
    e &\sim \mathrm{Rayleigh}(\sigma_{\rm R}=0.07).
\end{align}
In physical units, the semi-major axis distribution is centered at 45 au, with a spread of $\sim 5$ au, consistent with Fig. 5 in \citet{Adams2014}. The initial eccentricity spread of $\sigma_{\rm R}=0.07$ and mean of $\sigma_{\rm R} \sqrt{\pi/2} \approx 0.088$ are also roughly consistent with observed Kuiper Belt objects \citep[see, e.g.,][]{Tremaine2023}.
Second, we consider a detached disk-like distribution (abbreviated as DDO-like), with initial conditions
\begin{align}
    \log a/\rm au &\sim \mathcal{N}(\mu_{\rm N}=1.8, \sigma_{\rm N}=0.1) \\
    e &\sim \mathrm{Rayleigh}(\sigma_{\rm R}=0.2).
\end{align}
Here, the mean initial semi-major axis is $\approx 70$ au, its spread is $\approx 10$ au, and the eccentricity distribution is considerably warmer than the KBO-like setup \citep[$\sigma_{\rm R} = 0.2$, informed by][]{morbidelli2004}. Lastly, we also consider a distribution that is similar to the inner-Oort cloud (abbreviated as IOC)
\begin{align}
    \log a/\rm au &\sim \mathcal{N}(\mu_{\rm N}=4, \sigma_{\rm N}=0.1) \\
    e^2 &\sim \mathrm{Uniform}(0,1),
\end{align}
where the eccentricity distribution is now thermal, consistent with previous simulations of Oort cloud objects \citep[e.g.,][]{Vokrouhlicky2019}.
Here, we chose the semi-major axis distribution to again be a log-normal distribution, peaking at $10^4$ au with a spread of $\sim 5\times 10^3$ au.
The initial conditions for these populations are shown as dashed lines in Figs. \ref{fig:dbadist} (for semi-major axis) and \ref{fig:dbedist} (for eccentricity).

Fig. \ref{fig:dbadist} shows a comparison of the final semi-major axis distribution of these various reservoirs of debris in solid lines against their initial distributions.
The core of the KBO distribution remains similar in shape to its initial condition, albeit with the mean shifting to a higher semi-major axis and adopting a much wider spread. At large $a$, however, a small subset of KBO objects have begun to form a power law tail.
A similar behavior can be seen for the DDO distribution, though the spread is more exaggerated and the tail population extends well into the realm of the initial IOC distribution.
These behaviors for the KBO and DDO distributions are largely similar to the circular-like distributions at comparable $a$ from Fig. \ref{fig:paparams} --- a consequence of the fact that the energy evolution is agnostic of the distinct eccentricity distributions.
The IOC population experiences the most significant change in its distribution, where we observe diffusion to lower semi-major axes, but also a strong $\propto a^{-2}$ power-law tail at higher semi-major axes, which is caused by being far into the kick-dominated regime. The lower peak indicates that many of the IOC objects have been unbound by the TP-AGB evolution, with only 4\% of the initial population remaining below $a=10^5$ au.

In Fig. \ref{fig:dbedist}, the respective initial (dashed) and final (solid) eccentricity distributions for these populations are shown. We find that the eccentricities for KBO and DDO populations attain a greater spread, indicating that the populations have been heated. That said, the DDO population in particular experiences a very modest change. This is due to the fact that the KBO and DDO populations are at largely similar semi-major axes, but the DDO population is more eccentric: eccentricity diffusion slows as the distribution approaches thermal, and because the DDO-like population is closer to thermal than the KBO-like population, its change is much more subtle. An exaggerated version of this result can be seen in the IOC-like eccentricity distribution. Already being thermal, the eccentricity diffusion term in \eqref{eq:fp} is annihilated and therefore the IOC population experiences no further evolution in eccentricity.

\begin{figure}[htbp]
\centering
\includegraphics[width=1.0\linewidth]{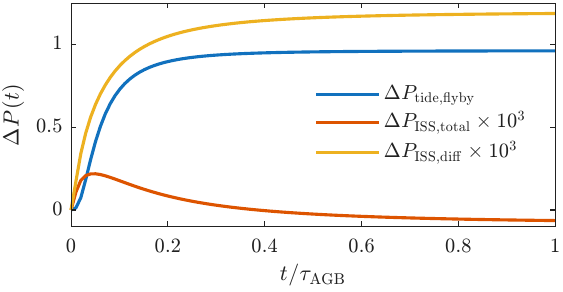}
\caption{Net change in probability density of IOC objects for a given region as a function of time. The ``tide,flyby'' term is the fraction of IOC objects pushed beyond $a=10^5$ au, which constitutes 96\% of the IOC population by the end of the TP-AGB phase. The $\Delta P_{\rm ISS}$ terms are integrated fluxes into the regions having perihelia within 10 au (inner solar system, abbreviated as ISS). By the end of the TP-AGB phase, the total ISS $\Delta P$ is negative, indicating that, overall, objects leave this region of the phase space. However, the diffusive contribution (``diff'') is positive (pushing objects \textit{into} the ISS), and dominates the total flux at early times.}
\label{fig:dbpflux}
\end{figure}
Of the populations we have considered here, the IOC-like population spans the largest range of distances from the TP-AGB star. Given that most of the IOC resides at very large semi-major axis and is therefore weakly bound, it is relatively easy to push these objects into the tide/flyby regime of Fig. \ref{fig:diag}, where they may eventually be unbound from the system entirely. On the other hand, some IOC objects are also on highly eccentric orbits, meaning that they are able to enter the inner solar system (abbreviated as ISS, which we delineate as $r_{\rm ISS} = 10$ au). Such objects can interact strongly with planets, which in turn can lead to their ejection from the system, or even their injection into the TP-AGB envelope. Understanding how these fringe populations evolve due to TP-AGB kicks is then of particular use to our understanding of the formation of interstellar objects, and perhaps even the pollution of WDs by metals. The rate at which these fringe populations evolve is given by the flux of probability
\begin{equation}
    q(t,E) = -A_E P(E) + \pD{E}{}\left(D_EP(E)\right)
\end{equation}
through the bound in question (tide/flyby or ISS). Note that, because the eccentricity distribution is already thermal, $P(E,\varepsilon)=P(E)$, and the $\varepsilon$ contribution to the flux is zero. In the case of the tide/flyby regime, this boundary is simply the value of $E$ corresponding to $a=10^5$ au (where the mass of the star is initially $M_\odot$, but decreases linearly to ${\approx}\,0.64\,M_\odot$). Therefore, the flux into the tide/flyby regime is
\begin{equation}
    q_{\rm tide,flyby}(t) = q\left(t,\,\frac{-\mu(t)}{2a_{\rm tide,flyby}}\right).
\end{equation}
For the ISS, we are concerned with the flux of probability into the region of $E-\varepsilon$ phase space where orbit \textit{perihelia} are located within $10$ au. Because perihelia vary with eccentricity, the semi-major axis boundary for the ISS is $a_{\rm ISS}(\varepsilon) = 10/(1-\varepsilon^{1/2})$ au. The total flux into the ISS is then the integral over the contributions at each $\varepsilon$:
\begin{equation}
    q_{\rm ISS}(t) = \int\limits_0^1\rmd\varepsilon \,  q\left(t,\,\frac{-\mu(t)}{2a_{\rm ISS}(\varepsilon)}\right).
\end{equation}
The evolution of the probability in these regions of interest is then given by the time integral  
\begin{equation}
    \Delta P(t) = \int\limits_0^t\rmd t' q(t').
\end{equation}
In Fig. \ref{fig:dbpflux}, we show the change in the probability of both the tide/flyby dominated and ISS regimes as a result of TP-AGB evolution, for the IOC-like distribution. Regarding the flux into the tide/flyby region of phase space, almost all IOC objects are ejected within a fraction of the AGB lifetime. The rapid onset of this flux is almost certainly a consequence of the establishment of the -2 power law tail to high semi-major axis. After the TP-AGB evolution is complete, 96\% of the objects in our IOC-like distribution occupy the regime in which forces from Galactic tides and stellar flybys are likely to unbind them from the resultant WD. We also provide the flux through the ISS boundary, including contributions from the advective and diffusive fluxes (``total''), as well as the contribution from the diffusive flux alone (``diff''). Considering all contributions to the flux, the fraction of IOC objects that transits the inner solar system experiences a net decrease by the end of the TP-AGB phase, a result that is driven by the advection due to mass loss and injection of energy by the kicks. That said, the diffusive flux, when considered by itself, drives a similar fraction of the population \textit{into} the inner solar system. In fact, at early times it dominates the total flux, and more IOC objects are pushed into the ISS than out of it. This implies that the competition between outward flux and inward flux in our IOC-like distribution is a close one: if the IOC population were initialized with a gradient that is slightly steeper in the vicinity of $a_{\rm ISS}$ than it currently is, the net change in ISS-transiting objects could easily become positive due to enhanced diffusion. We therefore cannot eliminate the possibility that, given a different, perhaps more realistic, initial distribution, TP-AGB kicks might instead increase the population of Oort cloud objects in the inner solar system.

\subsection{Evolution of Stellar Binaries}\label{sec:bnry}

We now investigate how distributions of stellar binaries are modified when TP-AGB kicks and mass loss are applied to one of the components. Given that $m_{\rm C}$ is now stellar, we no longer have that $m_{\rm AGB} \gg m_{\rm C}$, as in \S\ref{sec:slr}. We therefore introduce the mass ratio $R=m_{\rm C}/m_{\rm AGB}$ to distinguish between evolution in systems with varying companion masses. Here, all companions are assumed to be MS stars, therefore $R$ is strictly less than one, as the TP-AGB star must be more massive in order to leave the main sequence before its companion. We do consider binaries in which both components become WDs in \pII. With the above assumptions, it is also implied that $f$ is no longer the fraction of mass lost by the TP-AGB star alone ($f_{\rm AGB}$), instead it is given by
\begin{equation}\label{eq:rat}
    f = \frac{f_{\rm AGB}}{R+1}.
\end{equation}
We also consider TP-AGB stars of varying mass in this section, so in each case $f_{\rm AGB}(m_{\rm AGB})$ is determined according to the prescription we describe at the start of \S\ref{sec:nsln}. Lastly, although we have employed a fiducial $v_{\rm rms}$ of $1$ km/s thus far, previous one-kick studies of stellar-binary distributions have found that $\vrms$ may fall into a range of values \citep[e.g., $0.25{-}4$ km/s;][]{hwang2025}. As a potential alternative, then, here we also present Fokker-Planck solutions obtained for $v_{\rm rms}=0.5$ km/s.

The initial distribution of stellar-binary orbital parameters is the same for each solution that we present: The initial energy distribution (and thus semi-major axis distribution) is set according to a log-normal distribution in orbital periods $T$, satisfying
\begin{equation}
    \log T/{\rm day} \sim \mathcal{N}(\mu_{\rm N} = 4.8, \sigma_{\rm N} = 2.3).
\end{equation}
This choice is equivalent to that made by \citet{ElBadry2018}, which in turn was motivated by the observations of \citet{duquennoy1991}. The initial eccentricity distribution $P_0(e)$ is initialized at each $E$ according to an arctangent distribution:
\begin{equation}
    P_0(e) = \left(\arctan \kappa - \frac{\ln (\kappa^2+1)}{2\kappa}\right)^{-1} \arctan(\kappa e),
\end{equation}
where $\kappa$ is a parameter for the flatness of the distribution (larger $\kappa$ being flatter) that we set to 20 for each solution. This distribution is used instead of, say, a uniform distribution in $e$~\citep[\textit{not} in $e^2$;][]{duchenne2013}, to enable the Legendre decomposition of the initial condition, which requires that there be no binaries exactly at $e=0$. Regardless, although $P_0(e{=}0)=0$, it rises rapidly and becomes flat thereafter, mimicking a uniform distribution in $e$ with a deficit at $e=0$ (which is not unlike the observed distributions in \citealt{duchenne2013}). 

\begin{figure}[htbp]
\centering
\includegraphics[width=0.97\linewidth]{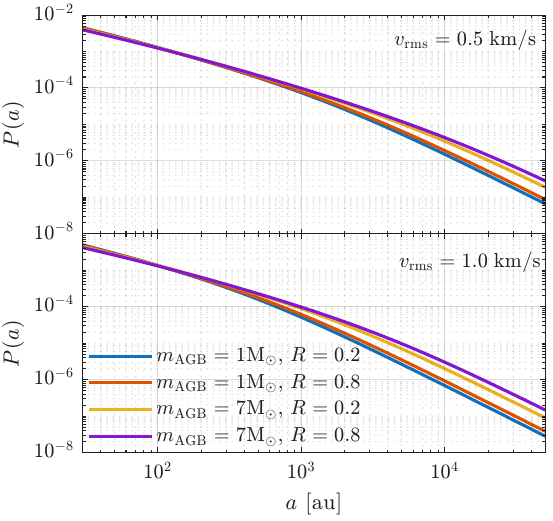}
\caption{Example post-TP-AGB $a$ distributions of stellar binaries having a variety of masses and mass ratios. The initial distributions in each case are those of \citet{duquennoy1991}. Each final distribution resembles a broken power law, reminiscent of the fits obtained by \citet{ElBadry2018}. We also include an additional RMS kick strength (0.5 km/s) to compare against results from our fiducial value (1 km/s).}
\label{fig:bina}
\end{figure}
Given the initial conditions and mass-loss prescription described above, in Fig. \ref{fig:bina} we show the post-kick $a$ distributions of binaries having $m_{\rm AGB}=1~\rm M_\odot$ and $7~\rm M_\odot$, and $R=0.2$ and $0.8$, as examples of how random TP-AGB kicks affect certain stellar populations. As stated earlier, we also include an alternate kick strength ($v_{\rm rms} {\approx}$0.5 km/s) to our fiducial value ($v_{\rm rms} {\approx} 1$ km/s) in order to explore the sensitivity of these distributions to the exact strength of TP-AGB kicks. In both panels, the region below $10^3$ au yields shallow slopes that reflect the (advected) initial lognormal distributions in orbital period $T$. This is the mass-loss dominated regime of the distribution evolution. Beyond, ${\sim}10^3$ au, however, the distributions steepen to the predicted $-2$ power law in semi-major axis due to kick diffusion, yielding an overall distribution that resembles a broken power law. Larger total binary masses experience this transition at slightly larger $a$, a consequence of the fact that for larger $\mu$, $\vwd$ represents a smaller fraction of the orbital energy. The comparison of the top and bottom panels demonstrates the reverse: that increasing the RMS kick strength from $0.5$ km/s to 1 km/s shifts the break in the distribution left by an amount that is roughly proportional to the change in the RMS kick energy. The qualitative properties of the distributions are unchanged, however.

\begin{figure}[htbp]
\centering
\includegraphics[width=1.0\linewidth]{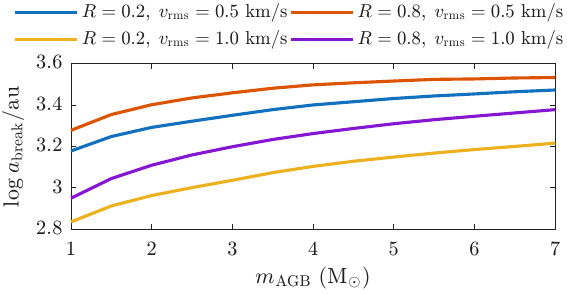}
\caption{Location of semi-major axis power law breaks, obtained by fitting broken power laws to the data in figure \ref{fig:bina}. The range of values shown here roughly coincides with the range obtained by \citet{ElBadry2018} for intrinsic separations ($10^{3.3} < s/{\rm au} < 10^{3.7}$).}
\label{fig:bbrk}
\end{figure}
To demonstrate how the break in the $a$ distribution depends on the TP-AGB mass and the mass ratio of the binary components, we fit broken power laws to $P(a)$ between $a = 30$ and $3\times 10^4$ au, and record the dependence of the break $a_{\rm break}$ as a function of $m_{\rm AGB}$. These results are shown in Fig. \ref{fig:bbrk} for $R=0.2$ and $0.8$, and $v_{\rm rms} = 0.5$ km/s and 1 km/s. Across the range of parameters shown, the logarithm of the break varies by less than an order of magnitude, spanning roughly $10^{2.8} < a/{\rm au} < 10^{3.6}$. By comparison, the broken-power-law fits that \citet{ElBadry2018} applied to the WD-MS intrinsic separation distribution found that the break most likely falls between $10^{3.3} < s/{\rm au} < 10^{3.7}$. This appears to be consistent with the range of $a_{\rm break}$ obtained from our Fokker-Planck solutions. Note that we do not attempt to predict the exact value of $a_{\rm break}$ for all binaries here because doing so requires binary population synthesis with an initial mass function, which this Fokker-Planck approach is not well-suited for (see \pII~for such a treatment).

\begin{figure}[htbp]
\centering
\includegraphics[width=1.0\linewidth]{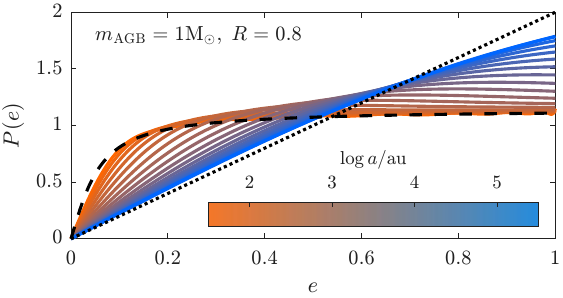}
\caption{Renormalized $a$-slices of the eccentricity distribution for the $m_{\rm AGB} = 1\rm M_\odot$, $R=0.8$ final binary population (solid), assuming $v_{\rm rms} = 1$ km/s. The initial distribution is given by the black dashed line, and the thermal distribution by the black dotted line. Eccentricity thermalization occurs rapidly around $a \approx 10^{3}$ au}
\label{fig:becc}
\end{figure}
As the location of $a_{\rm break}$ indicates the onset of the kick-dominated regime, it should also approximate the semi-major axis at which the eccentricity distribution becomes thermalized. In Fig. \ref{fig:becc}, the renormalized eccentricity distributions are shown for the $m_{\rm AGB} = 1~\rm M_\odot$, $R=0.8$ solution, as obtained from slices of $P(a,e)$ covering a range of $a$. The initial eccentricity distribution is shown as the black dashed line. Below $a=10^{2.5}$ au, the distributions change little, indicated by the tight packing of the orange curves. As $a$ approaches ${\approx}10^{3}$  au, the distributions begin to change rapidly, and approach the thermal distribution (the dotted black line). Beyond $a\approx 10^4$ au, the distributions change more gradually, now indicated by the tight packing of the blue curves. Comparing this result with Fig. \ref{fig:bbrk}, it is apparent that the semi-major axis at which the binary distribution becomes thermalized agrees with $a_{\rm break}$.

\begin{figure}[htbp]
\centering
\includegraphics[width=1.0\linewidth]{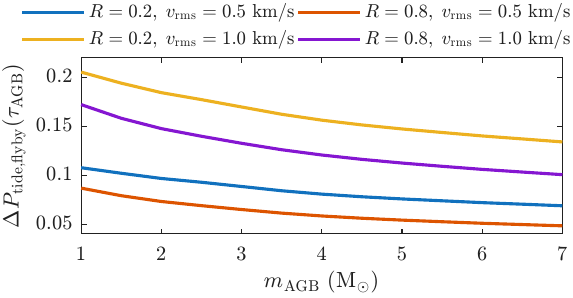}
\caption{Total change in probability density of stellar binaries in the Galactic-tide- and stellar-flyby-susceptible regime ($a>10^5$, where dissolution is implied), as a function of $m_{\rm AGB}$. Although dissolution does decrease with increasing binary mass, a factor of 10 change in binary mass only modifies $\Delta P(\tagb)$ by a factor of 2. Note that unlike \citet{hwang2025}, this fraction is normalized to the total population, not just those binaries between $a=10^3$ and $a=10^4$ au.}
\label{fig:btideq}
\end{figure}
As in the case of the inner Oort cloud, we expect some of the stellar binaries to become unbound after evolution as a result of being pushed into the regime in which Galactic tides and stellar flybys provide strong perturbations. In Fig. \ref{fig:btideq} we show the total flux of probability $\Delta P(\tagb)$ through the $a_{\rm tide,flyby}$ ($=10^5$ au) threshold as a function of the TP-AGB mass, for $R=0.2$ and 0.8, and $v_{\rm rms}=0.5$ and 1 km/s. Unsurprisingly, the populations that are most easily unbound are the lowest mass, with the $m_{\rm AGB} = 1~\rm M_\odot$ and $R=0.2$ population losing $22\%$ of binaries for $v_{\rm rms}= 1$ km/s. On the other hand, the $m_{\rm AGB} = 7~\rm M_\odot$ and $R=0.8$ population still loses $10\%$ of binaries for the same kick strength. Considering that the ratio of the total binary mass between those two populations is over a factor of ten, the fact that their $\Delta P(\tagb)$ differ by only a factor of two indicates that this flux is not particularly sensitive to the kick strength. This is likely a consequence of the flux being dominated by the power-law tail formation, which happens for any $v_{\rm rms}$ so long as it is strong enough to ensure $a_{\rm break}<a_{\rm tide,flyby}$.

\section{Discussion}\label{sec:disc}

\subsection{Comparison with Previous Studies of White-Dwarf Natal Kicks}

As discussed in \S\ref{sec:intro}, \citet{ElBadry2018} were the first to establish the connection between white-dwarf natal kicks and stellar-binary orbital-parameter distributions. In their work, only a single natal kick was assumed to occur to each TP-AGB star, and it was furthermore considered to be instantaneous (along with the associated mass loss). The intrinsic separation distributions for white-dwarf containing binaries were then obtained with the use of a numerical binary stellar evolution code. Although the simulated distributions resembled the observed distributions well enough to obtain an estimate of the RMS natal kick strength (${\sim}$0.9 km/s), the simulated distributions appeared \textit{qualitatively} distinct from the observed ones. The main feature was the lack of a power-law break---separation distributions obtained from the single-kick calculations were rounded, not clearly producing the power law tail expected from their observational constraints. A significant aspect of the random-walk approach to white-dwarf natal kicks is therefore that it does produce this broken power law, and that it does so with a $P(a)\propto a^{-2}$ tail, resembling the observed distributions.

\citet{hwang2025} extended the work of \citet{ElBadry2018} to focus on the eccentricities of these wide binaries. A pertinent result of theirs to our current investigation is that they found the eccentricity distribution of white-dwarf wide binaries to approach a thermal distribution around $10^3$ au. This agrees well with our random-walk approach. Although not presented here, in \pII~we show that at extremely large separations \citep[exceeding those simulated in][]{hwang2025}, singular (strong) natal kicks lead to a different result for the eccentricities: mildly super-thermal distributions. Because the single-kick paradigm entails much stronger individual kicks than our random-walk approach, kicks of any alignment relative to orbital motion are much more likely to increase the eccentricity. Thus, super-thermal eccentricity distributions can result. We note however, that while the results presented in \citet{hwang2025} show the eccentricity distributions approaching a thermal distribution, the possibility remains that at even larger separations they become super-thermal. Whether such super-thermal distributions are realized or not is also likely to have significant implications for the rates of tidal capture in white-dwarf binaries, a topic explored in depth by \citet{oconnor2026}.

During the preparation of this manuscript, we became aware of the concurrent work of \citet{Fuller2026}, which too assumes that AGB evolution induces a random walk in the velocity of white-dwarf progenitors. We briefly discuss here some differences that exist between our work and theirs. In their work, the random walk is not limited to the TP-AGB phase, but rather occurs over the entire AGB phase, which is much longer than the TP-AGB phase alone. All of the results we have presented here, however, evolve in normalized units of $t/\tagb$, thus our conclusions are not expected to be impacted by a change in this timescale (here $\tagb$ has been used as the TP-AGB lifetime for brevity, not the full AGB lifetime). The authors also take a different approach to determining the strength of the kicks than we do (in \pII): they estimate the recoil directly from a hydrodynamic model of the red-giant mass-ejection process~\citep{fuller2024}. Notably, the values they recover for the total kick strength are in approximate agreement with the results of previous studies and our own findings in \pII. The most significant difference between our work and theirs, however, is in the mathematical treatment of the evolving orbital parameter distributions. For example, \citet{Fuller2026} asserts that the diffusion of energy leads to the formation of a Gaussian probability distribution. However, because the diffusion coefficient depends on the energy, we do not find the result of our Fokker-Planck evolution to be a pure Gaussian. Instead, from our series solution \eqref{eq:fpesln}, we find that the distribution is driven toward to an exponential in energy (although the initial condition is often at least partially preserved, given finite $\tagb$). The same appears to be true for the comparison of our eccentricity distribution treatment with theirs, except that the eventual distribution is a thermal one.

\subsection{Future Work}\label{sec:future}

This work is intended as the first in a pair of two papers: here, we have focused on the theoretical aspects of how TP-AGB evolution affects the orbital-parameter distributions of WD binaries, planets, and debris disks. In \pII, we will present a detailed comparison between our model and observed white-dwarf binary distributions, so that constraints may be placed on both the strength of the individual kicks \textit{and} their quantity. Unlike the illustrative example distributions considered in this work, in \pII~we perform binary stellar evolution simulations given realistic initial conditions ---employing accurate initial mass functions and evolving binaries into the WD-WD phase as well. We furthermore include a direct comparison between all models proposed thus far: single instantaneous kicks, single gradual kicks, and many small kicks (in two and three dimensions). In combination, then, we offer a thorough analysis of the problem of random-walking WD natal kicks for stellar binaries that spans theoretical, numerical, and observational perspectives.

Given the results of \S\ref{sec:slr}, it is apparent that various systems beyond stellar binaries will be impacted by random-walking WD natal kicks.
As our own star will eventually enter a TP-AGB phase of its own, the fate of the solar system is closely tied to the strength of natal kicks and the associated mass loss. 
Because all planets in the solar system are strongly bound, the most immediate change in planetary orbits is adiabatic expansion due to stellar mass loss \citep[and the destruction of two or three planets;][]{sackman1993,esseldeurs2026}. 
That said, the random nature of the kicks could seed instability that develops post-TP-AGB evolution.
A future study may therefore investigate how solar-system planets evolve with post-main-sequence kicks and the subsequent secular evolution of these planets in the solar-WD system.
Further out, the Oort cloud likewise requires additional consideration as it lies at (and beyond) the junction between adiabatic and diffusive evolution. Interaction of kick-induced dynamics with additional forces from the Galactic tide and stellar flybys is therefore an intriguing future application of this work~\citep{Heisler1986, modak2023}.
The results of such a study would also inform the production rate of interstellar objects from evolved planetary systems.
Due to stellar evolution, these bodies can have different chemical composition compared to those ejected from newly formed planetary systems \citep{Katz2018,Pham2025}.

Beyond the fate of planetary systems, TP-AGB kicks have observational consequences tied to the emission from WDs themselves.
For instance, up to 50\% of WDs are observed to be ``polluted'' with metals in their atmosphere \citep[e.g.,][]{Koester2014, OuldRouis2024}.
Previous works suggested that this phenomenon can be explained through the delivery of evolved exoplanets, exomoons, exo-Kuiper Belts, exo-Scattered-disks, and exo-Oort clouds into WDs \citep[e.g.,][]{Alcock1986, Debes2002, Smallwood2018, Trierweiler2022, OConnor2022, OConnor2023, Pham2024, Veras2024}.
The random post-main sequence kicks described here can act as a universal mechanism to drive planetary bodies to high eccentricities, resulting in WD pollution.

Finally, we note that white-dwarf natal kicks could hold consequences for the formation of tight WD-WD binaries. Such binaries are expected to dominate the low-frequency ($<\rm mHz$) gravitational-wave signal of the upcoming Laser Interferometer Space Antenna \citep[LISA, cf.][]{Kremer2026}, forming a background that must be removed if weaker signals are to be explored at low frequencies. Tight-WD binaries are thought to result from common-envelope evolution, meaning that the stars must either be born sufficiently close together for CE evolution to occur on its own, or that some mechanism must push non-CE binaries into the CE regime. \citet{oconnor2026} suggested that recoil from AGB mass loss could drive binaries into this regime by way of an intermediate tidal-capture process. Determining whether a random-walk natal kick also supports this possibility is therefore another application of this work that should be explored. Unfortunately, quantitatively estimating the rate of binaries pushed into the CE regime requires a binary population-synthesis model that incorporates the initial mass function of stars and has the ability to resolve very high eccentricities.
Inquiry into this particular application is therefore left to \pII, where the aforementioned requirements are met.

\section{Conclusion}\label{sec:conclusion}

In this work, we modeled the orbital consequences of white-dwarf natal kicks, assuming that they are the result of an episodic mass-loss process that manifests as a random walk in the velocity space of the white-dwarf progenitor (a TP-AGB star). We derived a Fokker-Planck equation for the evolution of a probability distribution of two-body (binary) orbital parameters in which one of the bodies is both losing mass and being perturbed frequently by many small kicks.

Analytic investigation of this evolution equation yields certain observationally relevant conclusions. First, the populations of binaries (whether they be star-star or star-test-mass) develop a tail in the semi-major-axis distribution that scales as $P(a)\propto a^{-2}$ in the high $a$ limit. Second, the onset of this power-law tail is given by the semi-major-axis at which the mean-square cumulative kick strength, $v_{\rm rms}^2$, is comparable to the orbital energy. Third, the onset of the power-law tail in $a$ coincides with the separation at which the eccentricity distribution becomes thermal, $P(e)=2e$, regardless of its initial distribution at that semi-major axis.

We provide numerical solutions to the Fokker-Planck equation in the $a{-}e$ phase space for a variety of distinct initial distributions that mimic populations of planets, debris disks (similar to Kuiper-belt, detached-disk, and inner-Oort-cloud objects), and stellar binaries. We use these numerical solutions to verify our analytical conclusions, as well as to explore the various consequences that TP-AGB stars may have for such systems. Among these consequences are the findings that TP-AGB evolution will likely lead to the ejection of $>95\%$ of inner-Oort-cloud objects, it \textit{may} enhance the number of Oort-cloud objects that enter the inner solar system (depending on the initial distribution assumed), and that it causes ${\approx}5$ to $10\%$ of stellar binaries in the mass range $m_{\rm total} \in[1.2,7.2]\times M_\odot$ to be dissolved. 

\begin{acknowledgments}
This work benefited greatly from conversations with Shaunak Modak, Hanno Rein, Robert Ewart, Thomas Foster, Matthew Kunz, Ann-Marie Madigan, Drake Miller and Thomas Tauris.
We thank Nadia Zakamska, Chris O'Connor, and Yanqin Wu for inspiring us to study white dwarf natal kicks.
Support for S.M. was provided by NASA Astrophysics Program grant 80NSSC22K0828. D.P. acknowledges support from the McCray Postdoctoral Fellowship at the University of Colorado Boulder.
\end{acknowledgments}

\appendix

\section{Deriving the Fokker-Planck Jump Moments}\label{app:jmom}

In this Appendix, we give a brief overview of how the advection and diffusion jump moments are derived for the specific orbital energy and angular momentum.

\subsection{Energy Jump Moments}

The specific orbital energy of the binary is
\begin{equation}
    E = -\frac{\mu}{2 a} = \frac{v^2}{2} - \frac{\mu}{r}.
\end{equation}
After an instantaneous velocity kick $\Delta v$, and a mass-loss increment $\Delta \mu$, the change in $E$ is
\begin{equation}
    \Delta E  = \frac{(\Delta v)^2}{2} - \frac{\Delta\mu}{r} + v\Delta v\cos\psi
\end{equation}
Here, $\psi$ is the angle between the kick vector and the orbital velocity vector at the time when the kick occurred. When squared (as is needed for the diffusion moment), this becomes
\begin{align}
    (\Delta E)^2 =&~ \frac{(\Delta v)^4}{4} - \frac{(\Delta v)^2\Delta\mu}{r} + \frac{(\Delta\mu)^2}{r^2} \nonumber\\
    &+ v(\Delta v)^3\cos\psi + 2\Delta v\Delta\mu\paren{\frac{v}{r}}\cos\psi \nonumber\\
    &+ v^2(\Delta v)^2\cos^2\psi.
\end{align}
Assuming isotropic kicks, the distribution of $\psi$ introduces a dependence on the number of dimensions in which the kicks occur.

As described in \S\ref{sec:jmom}, the next step in the derivation is to average over the direction of the kicks, which, for constant kick strengths translates to an average over the angle $\psi$. For the advection moment $\Delta E$, this amounts to eliminating the term proportional to $\cos \psi$:
\begin{equation}
    \overline{(\Delta E)}=\frac{(\Delta v)^2}{2} - \frac{\Delta \mu}{r}.
\end{equation}
Assuming that the region of phase space that we evolve with this moment contains many binaries at unique orbital phases, our next step is to orbit average. This is done assuming that the binaries are uniformly distributed in the mean anomaly, and the combined kick-and-orbit average is denoted by $\oavg{\bcdot}$.
The orbit average of $1/r$ is well-known \citep[cf. Eq. 1.65 in][]{Tremaine2023}, yielding the result in \eqref{eq:eajmp}:
\begin{equation}
    \oavg{\Delta E} = -\frac{\Delta \mu}{a} + \frac{(\Delta v)^2}{2} = 2 E \paren{\frac{\Delta \mu}{\mu}} + \frac{(\Delta v)^2}{2}.
\end{equation}

The same two-step procedure can be done for $(\Delta E)^2$, where we now recognize that $\overline{\cos^2 \psi} = 1/d$, where $d$ is the number of dimensions the kicks occur in.
The kick average therefore yields
\begin{multline}
    \overline{(\Delta E)^2} = (\Delta v)^2\paren{\frac{v^2}{d}} +\frac{(\Delta\mu)^2}{r^2} - \frac{{\Delta\mu} (\Delta v)^2}{r} \\+ \frac{(\Delta v)^4}{4},
\end{multline}
and the subsequent orbit average (applied to $1/r$ and $1/r^2$) gives
\begin{multline}
    \oavg{(\Delta E)^2} = -2E\paren{\frac{(\Delta v)^2}{d}} + \frac{4E^2}{\sqrt{1-\varepsilon}} \paren{\frac{(\Delta \mu)^2}{\mu^2}} \\ 
    + 2\frac{E}{\mu}(\Delta v)^2\Delta \mu + \frac{(\Delta v)^4}{4}.
\end{multline}
Substituting $d=3$ and taking $\Delta \mu = 0$, we recover the result of \citet{hamilton2024b} for a binary experiencing three-dimensional impulsive kicks.

\subsection{Angular Momentum Jump Moments}

The specific orbital angular momentum $J$ relates to the eccentricity by
\begin{equation}
    \varepsilon \equiv e^2 = 1+\frac{2EJ^2}{\mu^2}.
\end{equation}
Due to the mass loss and kicks, we have changes in $E \to E+\Delta E$, $J \to J+\Delta J$, $\mu \to \mu+\Delta\mu$.
The change in $e^2 \equiv \epsilon$ is therefore
\begin{equation}
    \Delta\varepsilon = -\frac{2EJ^2}{\mu^2} + \frac{2(J+\Delta J)^2 (\Delta E + E)}{(\Delta \mu+\mu)^2}.
\end{equation}
This is already much more complicated than the $E$ moment, so we will have to make use of the $N\gg1$ assumption for the derivation to be tractable.
In the limit where the mass loss accompanying each individual kick is small ($\Delta \mu \ll \mu$), we can expand
\begin{align}\label{eq:deps}
    \Delta\varepsilon = \frac{2J^2E}{\mu^2}\Biggr(&\frac{2\Delta J}{J} + \frac{\Delta E}{E} - \frac{2\Delta\mu}{\mu} + \frac{(\Delta J)^2}{J^2} \nonumber\\
    &+ \frac{2\Delta J \Delta E}{J E} + \frac{3(\Delta\mu^2)}{\mu^2} - \frac{4\Delta J \Delta\mu}{J\mu} \nonumber\\
    &- \frac{2\Delta E\Delta\mu}{E\mu} +\mathcal{O}\paren{\varepsilon\frac{\Delta \mu^3}{\mu^3}}~ \Biggr).
\end{align}
We have previously found $\Delta E$, so we must now find the change in angular momentum $\Delta J$.

The change in the angular momentum vector due to the velocity kick is $\Delta\bb{J} = \bb{r} \cross \Delta \bb{ v}$ and we can decompose the $\Delta\bb{v}$ kick vector into the radial, transverse, and normal components as 
\begin{equation}
    \Delta\bb{v} = \Delta v(\sin\phi \sin\theta  \hat{\bb{r}} +  \sin\phi\cos\theta\hat{\bb{\theta}} + \cos\phi\hat{\bb{z}})
\end{equation}
where $\hat{\bb{r}}$ is parallel to the orbital radial vector, $\hat{\bb{\theta}}$ is perpendicular to $\hat{\bb{r}}$ in the orbital plane, and $\hat{\bb{z}}$ is parallel to the original angular momentum vector.
The angles $\theta \in [0,2\pi]$ and $\phi \in [0,\pi]$ are the azimuthal and polar angles of the kick, relative to the point in the orbit when the kick is applied.
An explicit decomposition in $(\phi, \theta)$ is needed in this case to expand the angular momentum vector (whereas $\psi$ is sufficient for the scalar quantity $E$).
Perturbing $J$, we find, with some manipulation:
\begin{equation}
    \Delta J = J \paren{\frac{k}{2} - \frac{k^2}{8} + \mathcal{O}(k^3)}
\end{equation}
where we have defined $k$
\begin{align}
    k &\equiv \frac{|\Delta \bb{J}|^2}{J^2} + 2 \frac{\bb{J}\bcdot\Delta\bb{J}}{J^2}.
\end{align}
Inserting our decomposition of the velocity vector, this equates to
\begin{align}
    k = \frac{r^2 (\Delta v)^2}{J^2} &\paren{\sin^2\phi \cos^2\theta + \cos^2 \phi} \nonumber\\
    &+ 2 \paren{\frac{r\Delta v\sin\phi\cos\theta}{J}}.
\end{align}
Note that $k$, being proportional to $\Delta v/v$, is a small quantity. In the expression for $\Delta J$, we can therefore choose to only keep terms up to second order in $k$:
\begin{equation}
    \Delta J = r \Delta v\sin\phi\cos\theta + r^2 \frac{(\Delta v)^2 \cos^2\phi}{2J}.
\end{equation}
This result, and the expression for $\Delta E$, can be substituted into the $\Delta \varepsilon$ expression (for brevity, we do not give that expression directly here).

Similar to the energy case, when $\Delta \varepsilon$ is determined in terms of $\Delta v$ and $\Delta \mu$, we first average over the kick directions by averaging over the angles $(\phi, \theta)$, and then average over orbital mean anomalies. The result, following some manipulation, is
\begin{align}
    \oavg{{\Delta\varepsilon}} =~ &(\Delta v)^2 \paren{\frac{-2-d+\varepsilon(7+d)}{2 d E}} \nonumber\\
    &+ (\Delta v)^2\frac{\Delta\mu}{\mu}  \frac{(1-\varepsilon)}{E} + \frac{(\Delta\mu)^2}{\mu^2}(1-\varepsilon).
\end{align}
The same procedure can be applied to $(\Delta \varepsilon)^2$, which yields
\begin{align}
    \oavg{(\Delta\varepsilon)^2} = &- (\Delta v)^2\frac{5 \varepsilon (1-\varepsilon)}{d E} \nonumber\\
    &- (\Delta v)^2\paren{\frac{\Delta\mu}{\mu}}(1-\varepsilon)^2\frac{2}{E}\paren{1+\frac{\varepsilon}{d(\varepsilon-1)}}.
\end{align}
Again, substituting $d=3$ and $\Delta \mu=0$ returns the equivalent expressions in \citet{hamilton2024b} for the case where there are three-dimensional velocity perturbations to the binary and no mass loss.

\subsection{The Cross-diffusion Moment}

The remaining jump moment to be determined is the cross-diffusion moment $\oavg{\Delta E\Delta \varepsilon}$. This is the most tedious of the moment calculations, so, for the sake of brevity we only demonstrate here that it contains no term of order $E\varepsilon \times N^0$.

The energy change $\Delta E$ contains terms that go like $\Delta v$, $\Delta \mu$, and $(\Delta v)^2$. Because the cross-diffusion moment entails multiplying these terms by $\Delta \varepsilon$, which itself contains terms that go like $\Delta v\sim v_{\rm rms}\sqrt{|\Delta \mu|/\mu}$ at lowest order, we need only consider the $v\Delta v \cos\psi$ contribution to $\Delta E$. All other terms in $\Delta E$, when multiplied into $\Delta \varepsilon$, will be higher order in $N^{-1}$. Similarly in \eqref{eq:deps}, only the first two terms in the parentheses, $2\Delta J/J$ and $\Delta E/E$, have contributions as low order as $\Delta v/v_{\rm rms}$. All of the other terms in parentheses, when multiplied by a term $\propto \Delta v$, will become higher order. The leading-order contribution to $\Delta E\Delta \varepsilon$ is then contained within
\begin{equation}
    \Delta E\Delta \varepsilon = v\Delta v \cos\psi \times \frac{2J^2E}{\mu^2}\left(\frac{2\Delta J}{J}+\frac{\Delta E}{E}\right). 
\end{equation}
Inserting the lowest-order contributions to $\Delta E$ and $\Delta J$, we have that
\begin{multline}\label{eq:cross1}
    \Delta E\Delta \varepsilon = (\Delta v)^2\frac{2J^2E}{\mu^2} \times \\ \biggl( \frac{2rv}{J}\cos\psi  \sin\phi\cos\theta 
    + \frac{v^2}{E} \cos^2\psi \biggr).
\end{multline}
It is now necessary to relate the angle used in the $E$ moments earlier ($\psi$) to the $\phi$ and $\theta$ employed in the determination of the $\varepsilon$ moments.

We can re-write $\cos\psi$ as
\begin{equation}
    \cos \psi = \frac{\bb{v}\bcdot\Delta\bb{v}}{v\Delta v} = \frac{v_r\Delta v_r + v_\theta \Delta v_\theta}{v\Delta v}.
\end{equation}
Using the fact the $\bb{J} \| \ez$, then the first term in the parentheses of \eqref{eq:cross1} becomes
\begin{equation}
    \frac{2rv\Delta v_\theta}{J\Delta v}\cos\psi= 2\left(\frac{v_r}{v_\theta}\sin^2\phi\sin\theta\cos\theta +\sin^2\phi\cos^2\theta\right)
\end{equation}
When averaging over the kick angles $\theta$ and $\phi$, the first term in parentheses above vanishes. So the kick-average of $\Delta E\Delta \varepsilon$ is, at this lowest order:
\begin{equation}\label{eq:cross2}
    \overline{\Delta E\Delta \varepsilon} = (\Delta v)^2\frac{2J^2E}{\mu^2} \biggl( \frac{2}{d} 
    + \frac{v^2}{Ed} \biggr).
\end{equation}
The final step is to orbit average, which only changes $v^2\rightarrow -2E$ in the second term in the parentheses above. The cross-diffusion jump moment $\oavg{\Delta E\Delta \varepsilon}$ is therefore 0 at order $E\varepsilon \times N^0$.

\section{Numerical Fokker-Planck Solver}\label{app:nums}

Here, we detail the numerical techniques used to obtain solutions to the $E{-}\varepsilon$ Fokker-Planck equation (\eqref{eq:fp}). As mentioned in \S\ref{sec:nsln}, our approach is based on the Legendre decomposition of the $\varepsilon$ domain. In this method, we first decompose the initial distribution to obtain the initial Legendre coefficients $F_{l,0}(E)$ of the expansion. Because all of the initial distributions in this work can be represented as $P_0(E,\varepsilon) = F_0(E)G_0(\varepsilon)$, the $F_{l,0}(E)$ are given by 
\begin{equation}
    F_{l,0}(E) = g_lF_0(E)
\end{equation}
where the $g_l$ are obtained through the Legendre polynomial orthogonality relationship as
\begin{equation}\label{eq:ortho}
    \quad g_l = \frac{2l+1}{2}\int\limits_{-1}^{1}p_l(\xi)G_0\left( \frac{1-\xi}{2}\right)d\xi.
\end{equation}
In the solutions of \S\ref{sec:nsln}, we use trapezoidal quadrature on a $\xi$ grid with a resolution of $10^3$ in order to numerically evaluate the integral in \eqref{eq:ortho}. For each initial eccentricity distribution, we calculate the first 100 Legendre polynomials this way, although the fact that high $l$ coefficients decay rapidly (\eqref{eq:lpoly}) means that we can often use as few as 20, so long as the initial eccentricity distribution $G_0(\varepsilon)$ is well-approximated by that many polynomials.

With the initial Legendre decomposition obtained, we may then evolve each $F_l(E)$ coefficient in time separately via \eqref{eq:lpoly}, as each coefficient is independent of the others. Because the initial conditions (and the solutions) span many orders of magnitude in $E$, we do not evolve the equation \eqref{eq:lpoly} directly in $E$, but rather do so on a logarithmic grid, in $x = \ln \left(-\frac{2df}{\vwd}E\right)$. In $x$, the evolution equation for each $F_l$ is given by
\begin{multline}\label{eq:logx}
    \pD{t}{F_l} = \frac{2}{\exp(x)}\pD{x^2}{^2F_l}+ \left[2-\frac{5df\,l(l+1)}{3\exp(x)}\right]F_l \\
    + \left[\frac{2\exp(x) + 2 + d}{\exp(x)}\right]\pD{x}{F_l}.
\end{multline}
The numerical solution to \eqref{eq:logx} is obtained using the method of lines, with central differencing employed for the $x$ derivatives. We employ the MATLAB stiff-ODE solver ode15s to advance the distribution forward in time, given the large variation in the coefficients present in \eqref{eq:logx}.

The resolution in $x$ for each solution obtained in \S\ref{sec:planets} is $10^3$, while the resolutions in \S\ref{sec:bnry} are $2\times 10^3$. At the lower $x$ boundary (which corresponds to the upper $E$ boundary) a rudimentary outflow boundary condition is employed, where the $x$ gradients in the ghost cells are equivalent to those at the edge of the domain. At the upper $x$ boundary (the lower $E$ boundary), the solution is set to $0$, which we ensure is already satisfied by the initial condition.

\begin{figure}[htbp]
\centering
\includegraphics[width=1.0\linewidth]{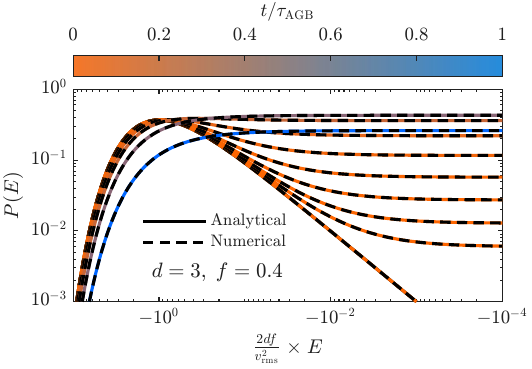}
\caption{Comparison of the numerical (dashed) and analytical (solid) solutions to \eqref{eq:fpe}, which are equivalent to the solution to \eqref{eq:lpoly} given $l=0$. Excellent agreement is obtained between the two approaches.}
\label{fig:numc}
\end{figure}
Although we do not have any analytic solutions in the full $E{-}\varepsilon$ phase space to directly compare our numerical solver to, the difference between the \eqref{eq:lpoly} and the energy-only equation \eqref{eq:fpe} is only a matter of an additional decay term, and we do have the ability to test our numerical solutions against exact solutions to \eqref{eq:fpe}. Fig. \ref{fig:numc} shows the time evolution of the analytic solution \eqref{eq:fpesln} (solid lines) compared to the numerical solution that we obtain for the same initial conditions (dashed) using the methods described here. Excellent agreement is obtained at all times. This is equivalent to the solution for the 0th Legendre coefficient, since there is no decay term. Higher Legendre terms are then evaluated with the decay term, and the full $E-\varepsilon$ solution is easily reconstructed using $P(E,\xi(\varepsilon)) = \sum_{l=0}^{n_l} F_l(E)p_l(\xi)$, where $n_l$ is the number of Legendre polynomials used.

\bibliography{references}{}
\bibliographystyle{aasjournalv7}

\end{document}